\documentclass[twocolumn]{aastex7}
\usepackage{mathtools}

\begin{document}

\title{Understanding the Origins of Super-Puff Planets: \\ A New Mass-Loss Regime Coupled to Planetary Evolution}

\author[0000-0003-3980-7808]{Yao Tang}
\affiliation{Department of Astronomy and Astrophysics, University of California, Santa Cruz \\
1156 High Street, Santa Cruz, CA 95064, USA}
\email{yaotang@ucsc.edu}

\author[0000-0002-9843-4354]{Jonathan J. Fortney}
\affiliation{Department of Astronomy and Astrophysics, University of California, Santa Cruz \\
1156 High Street, Santa Cruz, CA 95064, USA}
\email{jfortney@ucsc.edu}

\author[0000-0001-5061-0462]{Ruth Murray-Clay}
\affiliation{Department of Astronomy and Astrophysics, University of California, Santa Cruz \\
1156 High Street, Santa Cruz, CA 95064, USA}
\email{rmc@ucsc.edu}

\author[0000-0002-7520-5663]{Madelyn Broome}
\affiliation{Department of Astronomy and Astrophysics, University of California, Santa Cruz \\
1156 High Street, Santa Cruz, CA 95064, USA}
\email{mabroome@ucsc.edu}

\begin{abstract}
Super-puffs are a class of low-mass, large-radius planets that have challenged planet formation and evolution models. Their high inferred H/He mass fractions, required to explain their physical sizes, would lead to rapid atmospheric escape, raising questions about their long-term retention. Recent modeling work indicates that low-mass planets typically require 50\% less H/He mass to match their observed radius, due to significant roles of the radiative atmosphere and interior heating from the rock/iron core. Here, through a new quantitative analysis of XUV-driven escape in sub-Neptunes, we find that previous studies overestimated mass loss, as scaling laws in low-gravity regimes deviate greatly from the widely used energy-limited regime. We define a new regime, thermal-energy-mediated photoevaporation (TEMP), in which thermal energy conversion critically sets the mass-loss rate.  These effects make super-puffs more resilient to mass loss than previously thought. We develop a coupled evolution model integrating this updated thermal evolution framework with a 1D hydrodynamic photoevaporation model. Applying this novel, joint model to observed super-puffs and young low-density planets, we find that their masses, radii and transit pressures align with predictions assuming either a clear or hazy atmosphere. This indicates that super-puffs have undergone a combination of boil-off and photoevaporative mass loss, with boil-off dominating the process. Our results indicate that low-density planets typically possess both a thick convective envelope and substantial radiative atmosphere, which contribute to their large radii. For this to occur, these planets must have intermediate masses of 5-10$M_\oplus$ and receive stellar insolation $\lesssim 30F_\oplus$, favoring FG-type stars over M-dwarfs. 
\end{abstract}


\section{Introduction} \label{sec:intro}
Thanks to the Kepler and TESS missions, thousands of transiting planets have been found, dramatically expanding our understanding of planetary systems beyond our own. Among these, close-in exoplanets (with orbital periods $\lesssim$100 days) with physical sizes between Earth and Neptune are the most common \citep{Borucki11,Batalha12}. Observations reveal a clear size-based division into two distinct populations \citep{Fulton18}, sub-Neptunes (2–4 $R_\oplus$) and super-Earths (1–1.5 $R_\oplus$), separated by a radius valley. Structure modeling of sub-Neptunes suggests that their large radii require a few percent of their total mass to reside in a gaseous convective H/He envelope \citep{Lopez14,Rogers15,Owen17}. In contrast, bulk density analyses indicate that super-Earths have an Earth-like rock-iron composition \citep{Zeng16}.

A widely discussed explanation for this dichotomy is that some sub-Neptunes lose their H/He envelopes due to atmospheric escape, evolving into super-Earths and forming the observed radius gap \citep{Lopez13,Owen13,Ginzburg18,Owen20b}. Since atmospheric escape is highly sensitive to radius contraction driven by interior cooling, coupled evolution models \citep{Nettelmann11, Lopez12, Owen13, Jin14, Howe15, Chen16, Owen17, Ginzburg18, Rogers20, Tang25} that integrate thermal evolution and atmospheric escape have been developed. While these models can reproduce a radius gap similar to that of observation, the mechanisms governing the retention and evolution of H/He mass in these planets remain unclear based on existing observations \citep{Loyd20,Rogers21}, making this a central topic in exoplanet research. 

Three distinct mass loss mechanisms have been widely discussed, each driven by different energy sources, for decades. Photoevaporation \citep{Lammer03,Baraffe04,Yelle04} is driven by the X-ray and EUV (XUV) radiation deposited in a planet’s upper atmosphere, which has been extensively studied in the context of super-Earths, sub-Neptunes and gas giants. Core-powered escape \citep{Ginzburg16,Ginzburg18,Gupta19}, despite its name, is primarily driven by the energy from a bolometrically heated radiative atmosphere and is critically limited by the available interior cooling energy, operating on Gyr timescales.  Boil-off \citep{Owenwu16,Ginzburg16}, triggered by disk dispersal, creates a mismatch between the interior pressure and the decreased confinement pressure, leading to a rapid Parker wind mass loss in a few Myr. 

Subsequent analyses have shown that the detailed behavior of all three mechanisms depends on the evolution of a planet's interior. For example, in their boil-off model, \citet{Owenwu16} assume that envelope re-equilibration after mass loss is driven by cooling and contraction of the planet's convective envelope.  A recent study by \citet{Tang24}, in contrast, demonstrates that re-equilibration of the convective envelope does not require additional energy and instead results from redistribution of thermal and gravitational potential energy.  This subtle difference implies that mass loss cannot be limited by energy transport in the convective region, reducing the energy bottleneck critical to models of core-powered mass loss and suggesting that the role of interior cooling in controlling the mass loss rate was greatly overestimated in previous core-powered mass loss modeling \citep{Ginzburg18}. Though more indirectly, photoevaporation is also strongly dependent on the interior evolution of a planet---the mass loss rate is strongly dependent on the radius at which ionizing photons are absorbed \citep[see, e.g.,][Appendix A]{RMC09}, and following the evolution of this radius over time has important implications for a planet's mass loss history \citep{Lopez12, Owen24}.  In this paper, we explore the atmospheric evolution of a sub-Neptune via boil-off/core-powered mass loss and photoevaporation using a hydrodynamic model that is fully coupled to the sub-Neptune's thermal evolution.

Observations have revealed a peculiar class of sub-Neptunes, known as super-puffs, characterized by extraordinarily low bulk densities due to their low masses ($\sim 5 M_\oplus$) and large radii (up to $10R_\oplus$). Structure modeling by \citet{Lopez14} suggests that these planets commonly contain at least 30\% of their mass in a gaseous envelope to account for their inflated radii. However, \citet{Wang19} argued that such a massive envelope, while explaining the observed size, would not survive boil-off-driven mass loss. While super-puffs challenge traditional evolution models, photochemical modeling of sub-Neptunes' radiative atmospheres \citep{Wang19, Gao20, Ohno21} offers a potential alternative explanation: photochemical hazes may efficiently form and be lofted to very high altitudes at nbar pressure levels, inflating the apparent transit radius and lowering the mass loss rate \citep{Lammer16}. However, this scenario has had limited success in explaining all observed super-puffs \citep{Lopez14,Wang19}. Additional complications arise from planet formation theory, which suggests that the high H/He mass fractions from previous studies (often exceeding 30\%) cannot be accreted in-situ at the planets' current orbital distances \citep[for plants with mass $\lesssim 6 M_\oplus$,][]{Lee16}. 

\citet{Tang25} developed a new sub-Neptune thermal evolution model that predicts a systematically lower H/He mass fraction required to match a given planetary radius across a wide mass range ($\lesssim 20 M_\oplus$). This could potentially help to explain the existence of super-puffs. Importantly, the reduced envelope mass is more consistent with in-situ formation models. Given super-puffs' relatively high inferred H/He mass fraction \citep{Hallatt22} and their susceptibility to atmospheric escape \citep{Wang19,Gao20}, their physical properties are likely shaped by significant mass loss. In this study, we propose that super-puffs initially form in situ with a substantial (usually $\sim 30\%$ of the rock/iron core mass) H/He envelope and subsequently undergo a combination of boil-off and photoevaporation. The initial envelope mass is self-consistently calculated from formation models and is notably smaller than the values suggested by \citet{Lopez14} and \citet{Lee16}.

To test this hypothesis, we develop a coupled evolution model that integrates these mass loss processes with our recently published \citep{Tang25} thermal evolution model. A widely adopted analytical framework for atmospheric escape, the energy-limited approximation, has been used in many previous planetary evolution studies \citep{Nettelmann11,Lopez12,Owen13,Jin14,Howe15,Chen16,Owen17,Gao20,Rogers20}, but it can introduce large uncertainties when compared with full hydrodynamical results \citep{Erkaev16,Cubillos17,Kubyshkina18b,Krenn21}.

Early hydrodynamic simulations of photoevaporation, while pioneering, were not fully coupled to sub-Neptune thermal evolution models. This omission is nontrivial because mass loss and thermal evolution are strongly coupled \citep{Owen24,Tang24}, and the mass loss rate is highly sensitive to atmospheric lower boundary conditions. Specifically, previous hydrodynamical works often computed mass loss rates by interpolating from pre-computed hydrodynamic grids \citep{Lammer16,Kubyshkina18}, rather than calculating them self-consistently at each evolution timestep. Moreover, they either adopted a tabulated radius–envelope mass fraction relation that ignored ongoing thermal evolution \citep{Kubyshkina18}, or used analytical fits from thermal evolution models without incorporating feedback between mass loss and thermal cooling \citep{Fernandez23}.

To overcome this, we introduce numerical hydrodynamics of photoevaporation into sub-Neptune interior and evolution models for the first time. This approach significantly improves the accuracy of mass loss predictions. In addition, this evolution framework enables a joint, self-consistent evaluation of boil-off and photoevapration, unlike previous evolution models that either neglected boil-off entirely \citep{Lopez12,Owen13,Howe15,Chen16} or adopted post-boil-off mass fractions from \citet{Ginzburg16} \citep{Rogers23}, which are suggested to substantially overestimate initial H/He mass fractions for low-mass planets (Figure 11 of \citet{Tang24}). Previous hydrodynamic escape studies have modeled both mechanisms but without coupling to thermal evolution \citep{Kubyshkina18b}.

We test the proposed super-puff formation hypothesis by applying our fully coupled evolution model to observed super-puffs. Additionally, we include several observationally identified young, low-density planets in this analysis, as their early evolutionary stages directly reflect post-boil-off evolution conditions (e.g. envelope mass fraction and interior thermal state), making them ideal candidates for validating our evolution model. The planetary parameters used are listed in Table \ref{table:para}. Our objectives are to:
\begin{itemize}
    \item Gain new insights into photoevaporation in low-gravity planets, which is found in Section \ref{res:photo}.
    \item Compare our results to previous work in Section \ref{disc:comp}.
    \item Validate our evolution model against observations by matching our results to clear and hazy atmospheres, in Section \ref{res:app}.
    \item Determine the relative contributions of boil-off and photoevaporation and investigate the structure and formation conditions of super-puffs, in Section \ref{res:implication}.
    \item Constrain the metallicity of super-puffs, in Section \ref{res:metallicity}.
\end{itemize}
Beyond this, Section \ref{sec:methods} introduces the key components of the new proposed evolution framework. 

\begin{deluxetable*}{c c c c c c c c c c}
\tabletypesize{\scriptsize}
\tablewidth{0pt} 
\tablecaption{Physical parameters for Super-puffs and young planets modeled}\label{table:para}
\tablehead{
 \colhead{Name} & \colhead{Mass ($M_\oplus$)} & \colhead{Radius ($R_\oplus$)}  & \colhead{$T_{\rm{eq}} (K)$} & \colhead{a (AU)} & \colhead{Age (Gyr)} & \colhead{$M_*$ ($M_\odot$)} & \colhead{$R_*$ ($R_\odot$)} & \colhead{$L_*$ $(L_\odot$)} & Reference
} 
\startdata 
Kepler-51b & $6.7_{-2.8}^{+2.8}$ & $6.83_{-0.13}^{0.13}$ & 512 & 0.249 & $0.5_{-0.3}^{+0.7}$ & 0.96 & 0.87 & 0.71 & 1\\ [0.5 ex]
- & $3.5_{-2.1}^{+2.1}$ & - & - & - & - & - & - & - & -\\ [0.5 ex]
Kepler-51c & $6.09_{-0.41}^{+0.41}$ & $6.4_{-1.4}^{+1.4}$ & 414 & 0.381 & - & - & - & - & -\\ [0.5 ex]
- & $5.65_{-0.81}^{+0.81}$ & - & - & - & - & - & - & - & -\\ [0.5 ex]
Kepler-51d & $6.6_{-1.0}^{+1.0}$ & $9.32_{-0.18}^{+0.18}$ & 359 & 0.505 & - & - & - & - & -\\ [0.5 ex]
- & $5.6_{-1.2}^{+1.2}$ & - & - & - & - & - & - & - & -\\ [0.5 ex]
Kepler-47c & $3.17_{-1.25}^{+2.18}$ & $4.65_{-0.07}^{+0.09}$ & 263 & 0.96 & $4.5_{-0.5}^{+0.5}$ & 0.96 & 0.94 & 0.77 & 2\\ [0.5 ex]
Kepler-79d & $11.3_{-2.2}^{+2.2}$ & $6.912_{-0.014}^{+0.014}$ & 654 & 0.287 & $3.24_{-1.07}^{+0.78}$ & 1.24 & 1.28 & 2.70 & 3\\ [0.5 ex]
- & $6.0_{-1.6}^{+2.1}$ & $7.16_{-0.16}^{+0.13}$ & - & - & - & 1.165 & 1.302 & - & 4\\ [0.5 ex]
Kepler-87c & $6.4_{-0.8}^{+0.8}$ & $6.14_{-0.29}^{+0.29}$ & 403 & 0.676 & $7.5_{-0.5}^{+0.5}$ & 1.1 & 1.82 & 2.92 & 5\\ [0.5 ex]
HIP 67522b & $13.8_{-1.0}^{+1.0}$ & $9.763_{-0.504}^{+0.493}$ & 1171 & 0.0747 & $0.017_{-0.002}^{+0.002}$ & 1.22 & 1.38 & 1.75 & 6,7\\ [0.5 ex]
V1298 Tau b & $8_{-2}^{+4}$ & $9.95_{-0.37}^{+0.37}$ & 685 & 0.172 & $0.025_{-0.005}^{+0.005}$ & 1.1 & 1.3 & 1 & 8\\ [0.5 ex]
- & $12_{-1}^{+1}$ & - & - & - &- & - & - & - & 9\\ [0.5 ex]
Kepler-223d & $8.0_{-1.3}^{+1.5}$ & $5.24_{-0.45}^{+0.26}$ & 797 & 0.122 & $4.27_{-2.48}^{+3.92}$ & 1.125 & 1.72 & 2.35 & 10\\ [0.5 ex]
Kepler-223e & $4.8_{-1.2}^{+1.4}$ & $4.60_{-0.41}^{+0.27}$ & 725 & 0.148 & - & - & - & - & -\\ [0.5 ex]
\enddata
\tablerefs{(1)\citet{Masuda24}; (2)\citet{Orosz19}; (3)\citet{Yoffe21}; (4)\citet{Jontof-Hutter14}; (5)\citet{ofir14} (6)\citet{Thao24}; (7) \citet{Rizzuto20}; (8)\citet{Barat24}; (9)Barat et al. (2025, in review)(10)\citet{Mills16}
}
\tablecomments{ A dash symbol indicates that the value in the cell is the same as in the row above. The system age for Kepler-51 is taken from \citet{Roberts20}, while the ages for Kepler-79 and Kepler-223 are from \citet{Morton16}. Stellar luminosities for Kepler-47, Kepler-79, and Kepler-223 are obtained from TICv8 \citep{Stassun19} and Gaia-based data \citep{Berger18}, as effective temperatures are not available in the cited references. For Kepler-87c, an Earth-like Bond albedo of 0.3 was assumed in the referenced study, otherwise it is assumed to be zero. The equilibrium temperature ($T_{\rm{eq}}$) and semi-major axis ($a$) are derived from stellar parameters. For each Kepler-51 planet, two mass estimates \citep{Masuda24} are provided. The first rows (with higher masses) assume an external 2:1 resonance between Kepler-51e and Kepler-51d in their TTV fit, while the second rows (with lower masses) assume an internal 2:1 resonance.
}
\end{deluxetable*}

\section{Model Description} \label{sec:methods}

\subsection{Interior Structure and Evolution}
\label{subsec:evolution}
Our structure and evolution modeling is based on the framework developed in \citet{Tang25}. This model computes the physical (radius-pressure) and thermal (temperature-pressure, hereafter $T$-$P$) structures of sub-Neptune planets. The interior consists of an iron core, a silicate mantle, and a convective H/He envelope, while the radiative atmosphere above the convective zone is computed up to nbar pressure levels. Additionally, it self-consistently accounts for solidification processes within the rock/iron core through parameterized convection, commonly used in terrestrial modeling. This approach models each interior layer with independently evolving adiabats and calculates the energy transport rate by solving non-adiabatic thermal boundaries. Specifically, latent heat release during solidification injects additional internal energy, sustaining a larger planetary radius and enhancing atmospheric escape efficiency. Tidal heating is not included in our interior model; we assess its impact \emph{post facto} in Section \ref{disc:tides}.

Each structural layer is assumed to be in hydrostatic equilibrium, except for the radiative atmosphere at young ages, which is subject to boil-off outflows. We model the boil-off phase using a steady-state, energy-limited isothermal Parker wind, as detailed in \citet{Tang24}. This phase determines the initial H/He mass fractions and interior thermal states for subsequent evolution. The initial envelope mass fraction for the boil-off phase is derived from the planet formation model of \citet{Ginzburg16} and is given by:
\begin{equation}
\label{initial}
f_{0} = 0.02(M_{\rm{ric}}/M_\oplus)^{0.8}(T_{\rm{eq}}/1000K)^{-0.25}(t_{\rm{disk}}/1 \mathrm{Myr})^{0.5}
\end{equation}
where $M_{\rm{ric}}$ is the rock/iron core mass, $T_{\rm{eq}}$ is the atmospheric equilibrium temperature and $t_{\rm{disk}}$ is the disk lifetime. Observationally, disk lifetimes are constrained to be 3–10 Myr \citep{Mamajek09, gorti2016}, with a recent study reporting a longer lifetime of 30 Myr \citep{Long2025}. To remain conservative, we adopt 10 Myr in this work. We have tested a shorter value of 5 Myr, as assumed in previous modeling efforts. It decreases the planetary radius by less than 5\% when the planetary mass is matched at the observed age, which does not qualitatively impact our results. For planets that undergo a boil-off phase, the subsequent evolution is not highly sensitive to $f_0$. The initial thermal state of the interior is set by equating $t_{\rm{disk}}$ with the envelope's Kelvin-Helmholtz contraction timescale while assuming thermal equilibrium between the core, mantle, and envelope.

Once the atmospheric blow-off phase ends after 1-3 Myrs, we hydrostatically integrate the entire planet, including the radiative atmosphere. At young ages, this can overestimate the planetary radius and is inconsistent with an isothermal Parker wind. We introduce an approach to smoothly transition between the two regimes in Section \ref{subsec:photo}. We are particularly interested in evaluating different pressure levels to match observed planetary radii. We adopt a reference pressure of 20 mbar for the optical transit radius of a clear atmosphere and nbar levels for hazy atmospheres. The evolution model improves upon previous sub-Neptune models by incorporating a more realistic $T$-$P$ profile from an analytical radiative transfer model of \citet{Guillot10} (which is hydrostatic and gray), vertically varying gravity, and a molecular-to-atomic transition in the radiative atmosphere due to photo- and thermal- dissociation. As a result, this model predicts larger planetary radii at a given envelope mass fraction, particularly at low-pressure levels, facilitating explanations for super-puff planets. However, this also enhances atmospheric mass loss due to the reduced gravitational binding.

We assume the metallicities of the convective envelope and radiative atmosphere are the same and choose one times solar as the default, unless otherwise specified in Section \ref{res:metallicity}. We find a negligible difference between a pure H/He envelope and a one-times solar metallicity envelope. The metallicity is modeled by incorporating a mass fraction $Z$ of water into the envelope. A metal-enhanced planet cools at a different rate and exhibits different atmospheric mean molecular/atomic weights. The implementation of this process is described in \citet{Tang25}, and its implementation in photoevaporation is discussed in Section \ref{subsec:photo}.

While our model can estimate the static structure of envelope-free terrestrial planets, it does not capture their evolution. 
Consequently, we exclude planets that evolve into envelope-free states from our study. This transformation can be driven by both mass loss \citep{Lopez12} and thermal evolution \citep{Tang25}; the latter occurs when the radiative-convective boundary (RCB) reaches the mantle-envelope boundary.

\subsection{Photoevaporation}
\label{subsec:photo}
A novel aspect of our work is the coupling of a one-dimensional hydrodynamic photoevaporation model with the evolution model, which we run at each evolution timestep. This model computes the mass loss rate by solving steady-state hydrodynamics and the ionization balance of the XUV-driven wind in the upper atmosphere using a relaxation method. It employs a simplified radiative transfer scheme for XUV fluxes and includes cooling mechanisms such as $PdV$ work, advective cooling, and Lyman-alpha (Ly-$\alpha$) cooling. Our model is based on the Broome et al. (2025, submitted) framework for ``typical-gravity'' giant planets (with masses $>20 M_\oplus$), incorporating both X-ray-driven escape and atomic metal species from the original hot-Jupiter model of \citet{RMC09}. Additionally, it models bolometric heating and cooling processes below the photoionization base. However, for low-gravity planets, modeling these processes at high pressure levels ($>1\mu$bar) becomes challenging due to uncertainties in the (assumed constant) opacities. We define the photoionization base as the location where ionization heating maintains the atmosphere at a temperature higher than the bolometric-driven skin temperature.

We have optimized the performance of the photoevaporation model to ensure compatibility with low-gravity planets. Compared to Saturn- to Jupiter-mass planets and terrestrial planets, low-gravity sub-Neptunes exhibit a larger atmospheric scale height $H$ relative to their envelope size, positioning the ionization base at a radius $R_{\rm{base}}$ readily exceeding twice the radiative-convective boundary radius $R_{\rm{RCB}}$. Their low escape velocity, $v_{\rm{esc}}=\sqrt{G M_p / R_{\rm{base}}}$, indicates that the physical structure below $R_{\rm{base}}$ is non-hydrostatic, even beyond the boil-off phase. In such cases, since the density falls off with altitude more rapidly in a steady-state wind than in a hydrostatic atmosphere, assuming a hydrostatic atmosphere can significantly overestimate $R_{\rm{base}}$, leading to physical inconsistencies. Specifically, if $R_{\rm{base}}$ were to be located above the isothermal sonic point $R_{\rm{s}}$ of the bolometrically heated outflow (set by $T_{\rm{eq}}$), the hydrostatic assumption would incorrectly suggest that the planet is still undergoing boil-off, when in reality, it has transitioned to photoevaporation. We note that in a wind undergoing boil-off, if $R_{\rm{base}}>R_{\rm{s}}$, photoionization in the supersonic region does not contribute additional mass loss, and photoevaporation does not occur under these conditions. The isothermal sonic point is given by:
\begin{equation}
\label{sonicpoint}
R_s = \frac{GM_p}{2c_s^2}
\end{equation}
where $c_s=\sqrt{kT_{\rm{eq}}/(\mu m_H)}$ is the sound speed of the isothermal Parker wind and $T_{\rm{eq}}=[F_{\rm{bol}}/(4\sigma)]^{1/4}$ is the equilibrium temperature. 

As a super-puff thermally evolves, its RCB radius shrinks and surface gravity increases. Consequently, the atmosphere below the photoionization base transitions from steady-state to hydrostatic. This transition is characterized by the wind velocity, which decreases monotonically with decreasing atmospheric scale height during post-boil-off evolution. The hydrostatic model represents the zero-velocity limit of the steady-state wind solution, a state achievable only when the external confinement pressure is sufficiently strong ($>p_{\infty} \equiv p_0\exp(-H/R_0)$, where $p_0$ and $R_0$ are the reference pressure and radius) to counterbalance the planet’s gravitational force. This confinement pressure likely originates from stellar winds in the post-formation phase and is typically on the order of picobars.

\begin{figure}
\centering
\includegraphics[width=0.45\textwidth]{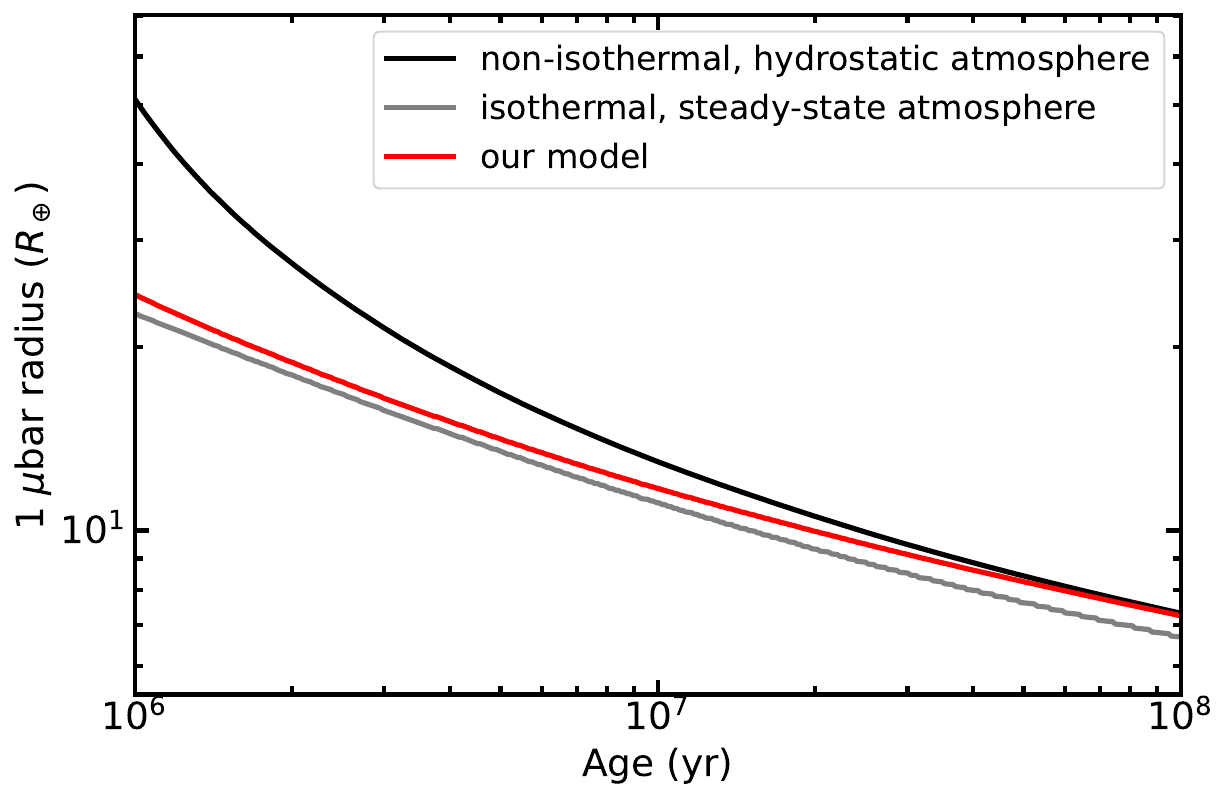} 
\caption{ 1$\mu$bar radii calculated from the isothermal Parker wind (gray) and our non-isothermal, hydrostatic atmosphere (black), which serves as the fiducial model for determining optical radii. The transition between these two regimes (red) ensures a proper estimate of the lower-boundary radius for the photoevaporation model. 
}
\label{transition}
\end{figure}

This transition provides the appropriate lower-boundary condition for the photoevaporation model by determining the location of the photoionization base, which is essential for coupling the hydrodynamic photoevaporation and evolution models. To accurately capture this transition, we implement a continuous transition between the two regimes. This modeled radius is shown in red in Figure \ref{transition}. We compare it to two limiting cases: the radii calculated from a non-isothermal hydrostatic model ($R_1$, black) and from an isothermal, steady-state wind model($R_2$, gray). Notably, these two curves do not naturally converge as the planet thermally contracts, since our model does not capture non-isothermal bolometric-driven winds. The transition is modeled using a hyperbolic function:
\begin{equation}
\label{activation}
\phi = \frac{1}{2} + \frac{1}{2}\tan\left[\frac{\log(\Tilde{v})-\log(\Tilde{v}_{\rm{crit}})}{w_{\rm{log,v}}}\right] \\
\end{equation}

\begin{equation}
\label{Rtrans}
R = \phi R_1 + (1-\phi)R_2
\end{equation}
where $\Tilde{v}$ is the wind velocity normalized by the isothermal sound speed $c_s$ (Mach number). Note that $\Tilde{v}$, $R_1$, $R_2$ and $R$ are defined at the same pressure level. \citet{Tang25} demonstrated that maintaining a relative radius difference below 5\% between the hydrostatic and steady-state models requires a critical Mach number of $\Tilde{v}_{\rm{crit}} \lesssim 10^{-8}$. We adopt a log transition width $w_{\rm{log,v}}=4$. The resulting radius transition between the two regimes (red) usually occurs around $\lesssim$10 Myr for most of the modeled planets. 

We set the lower boundary of the photoevaporation model at 1$\mu$bar, using the radiative atmosphere model described in Section \ref{subsec:evolution} and Equations \ref{activation} and \ref{Rtrans} to provide the lower boundary conditions, including temperature, radius, and density. In other words, we do not explicitly compute the wind structure dominated by bolometric radiation in the lower atmosphere down to the RCB in the photoevaporation model, as this omission improves the model's numerical convergence. While $PdV$ work in the lower atmosphere cools the wind to temperatures lower than those predicted by the hydrostatic radiative-transfer model, Figure 5 of \citet{Tang24} shows that this bolometric-driven wind does not become critically bolometric-limited after boil-off, making our treatment a reasonable approximation. Addressing this effect could be a focus of future work. Within the parameter space explored, no planet has an XUV photoionization base below the 1$\mu$bar level.

Photoevaporation begins when the photoionization heat flux penetrates below the isothermal sonic point of the bolometric-driven Parker wind \citep{Owen24}. This is expected because EUV photons have a significantly larger cross-section compared to the re-emitted thermal radiation from deeper atmospheric layers, the dominant heat source for the bolometric-driven wind at $R_s$. As a result, even a small fraction of the incident XUV flux reaching below the sonic point can heat the atmosphere beyond the skin temperature, initiating photoevaporation. We find that this mechanism holds even for low-gravity planets, although their extended atmospheres with a large scale height distribute the photoionization heating over a larger thickness, leading to a lower energy input per unit volume; in contrast, for giant planets, photoionization heating is deposited in a much thinner layer near the photoionization base, resulting in a more localized and intense heating effect. 

The mass loss transition time can be approximated as the time at which the optical depth to EUV photons at the sonic point, $\tau_{\rm{EUV,s}}$, reaches unity. This optical depth is obtained by integrating the opacity of an isothermal Parker wind from infinity. To first order, this integral simplifies to:
\begin{equation}
\label{taus}
\tau_{\rm{EUV,s}} = \sigma_{EUV} H_s n_s
\end{equation}
where $\sigma_{EUV}$ is the absorption cross-section of EUV photons, and $H_s$ and $n_s$ are the atmospheric scale height and number density at $R_s$, respectively. In this calculation, we only considered EUV photons rather than X-ray, as we find they dominate the photoevaporation from low-gravity planets (see below). The transition to photoevaporation typically occurs a few million years after the end of boil-off, which we define as the point when the mass loss timescale of the isothermal Parker wind exceeds the thermal contraction timescale. Since the total mass loss from the Parker wind during this transitional period is negligible, the end of boil-off and the onset of photoevaporation can be considered effectively simultaneous \citep{Tang24}.

The photoionization model accounts for the spectral distribution of XUV photons to determine the energy available at each wavelength. Since this study focuses on super-puffs, which have only been found around G-type and late F-type stars, we assume a Sun-like XUV spectrum as a reasonable approximation.

We also incorporate metallicity in the photoionization model. Although the photoionization model is capable of capturing the ionization balance and cooling effects of atomic metals, we exclude these processes for computational efficiency. For low-gravity planets, we find that the ionization fraction and electron number density remain low throughout most of the wind, leading to inefficient metal line cooling (see Section \ref{res:photo}). As a result, we model only hydrogen and helium, adjusting their mass fractions to match the mean molecular weight at a given metallicity. This adjustment is essential for setting the location of $R_{\rm{base}}$ in the hydrodynamic calculations. We further justify this simplification in the Appendix \ref{app:mmw}. The default helium mass fraction, Y=0.275 at zero metallicity, is consistent with that used in the evolution model. The impact of mean molecular weight on photoevaporation is discussed further in Section \ref{res:metallicity}.

\subsection{Stellar Evolution}
\label{subsec:stellar}
We model stellar evolution to account for the time-dependent XUV and bolometric fluxes, which are crucial for determining the mass loss rates from photoevaporation and boil-off, respectively. Empirically, the XUV flux also depends on stellar rotation \citep{Tu2015,Johnstone21}, as the rotation rate strongly influences magnetic activity, which in turn heats the outer stellar atmosphere and drives XUV emission.

Both X-ray and EUV luminosities remain saturated above a certain rotation threshold, after which they decay over time as stellar rotation slows due to magnetic braking. The saturation level and decay rate vary across different spectral types, primarily as a function of stellar mass. We adopt the EUV flux evolution (in units of $erg/s/cm^2$), which corresponds to moderately rotating G-type stars as given by \citet{Ribas05}:
\begin{equation}
\label{EUV}
F_{\rm{EUV}} = 29.7*\max{(t,0.1)}^{-1.23}*a^{-2}
\end{equation}
where $t$ is the stellar age in Gyr and $a$ is the semi-major axis in AU. The original work derives this relationship by scaling the stellar radius to the solar radius at the sample star’s age. Here, for simplicity, we further scale Eq. \ref{EUV} by the stellar bolometric luminosity, which is reasonable given that the host stars considered in this study have similar effective temperatures.

For the only non-G-type star in our sample, Kepler-79 (a late F-type star), we model the X-ray luminosity using the empirical scaling relations from \citet{Jackson12}, as direct EUV flux measurements for this spectral type are less readily available.

These scaling relations generally predict that later-type stars emit a lower fraction of their radiation in the X-ray bands relative to their bolometric luminosity. To estimate the EUV luminosity, we use the X-ray luminosity and apply the scaling from \citet{Sanz-Forcada11}, which is based on extrapolated EUV flux from coronal models. However, this method introduces greater uncertainty for Kepler-79 compared to G-type stars. Theoretical models exist for rotational and XUV evolution \citep{Johnstone21}, but they depend on the unknown initial stellar rotation rate. Constraining this parameter is beyond the scope of our study, and the relative magnitude of uncertainties in the theoretical versus observational approaches remains unclear. For simplicity and to reduce model dimension, we therefore adopt the observation-based method of \citet{Sanz-Forcada11}. Finally, we determine the energy density at each wavelength by scaling the solar XUV spectrum to match the integrated EUV flux derived above.

The intensity of boil-off is highly sensitive to equilibrium temperature \citep{Tang24}. To accurately model equilibrium temperatures at the time of boil-off, we incorporate a stellar bolometric luminosity model based on \citet{Spada13}. Specifically, we compute the average bolometric luminosity over the 10–30 Myr age range, during which boil-off is likely to occur. We find that F- and G-type stars have already exited their initial contraction phase and are beginning their main sequence evolution by this time. Consequently, their bolometric luminosities in the boil-off phase are approximately 0.5–0.8 times the present-day stellar luminosity for Gyr-old stars, as they have not yet undergone significant radius inflation. For young stars ($<1$Gyr), this bolometric luminosity evolution is negligible.

Additionally, we track the total pressure of stellar winds throughout all evolutionary stages using the stellar wind model from \citet{Johnstone15}. For close-in planets at young ages, strong stellar winds may suppress EUV-driven mass loss by confining the hydrodynamic outflow. 
This effect may be particularly significant for super-puffs, whose large scale heights place the photoionization base at low-pressure regions. We explore this possibility further in Section \ref{disc:wind}.

\section{Results} \label{sec:results}
\subsection{Photoevaporation in Low-Gravity Planets}
\label{res:photo}
In this section, we examine photoevaporation in low-gravity planets. Traditionally, the energy-limited mass loss prescription \citep{Watson81,Erkaev07}, developed for photoevaporation in hot-Jupiters, has been widely adopted to estimate photoevaporative mass loss in sub-Neptune modeling \citep{Nettelmann11,Lopez12,Owen13,Jin14,Howe15,Chen16,Owen17,Gao20,Rogers20}. In this section, we reassess this treatment. We demonstrate that sub-Neptunes' mass loss characteristics diverge significantly from those of high-gravity planets in Section \ref{subsec:comparison}. We propose a comprehensive framework for classifying photoevaporation regimes based on planetary parameters in Section \ref{subsec:scaling}. 

\subsubsection{Super-Puffs vs. Hot-Jupiters} \label{subsec:comparison}
Figure \ref{structure} compares the wind structures of a super-puff (black curves), represented by Kepler-51b at the onset of its photoevaporation phase (a few Myr), and a hot Jupiter (red curves), modeled after HD 209458b at its current age. The hot-Jupiter receives approximately five times more EUV flux than the super-puff, due to its closer orbital distance. In the radiative atmosphere below $R_{\rm{base}}$, the gravity of HD 209458b remains Earth-like ($1g$), whereas Kepler-51b exhibits significantly lower gravity-—only $0.016g$ at $R_{\rm{base}}$ and $0.126g$ at the RCB. The panels from the top left to bottom right in Figure \ref{structure} present the wind temperature, Mach number, density, and ionization fraction, with the x-axis normalized by the optical transit radius $R_p$, assuming a clear atmosphere.

In the top left panel, the photoionization bases are marked by circles. For super-puffs, $R_{\rm{base}}$ is located at a considerable distance from both $R_p$ and the envelope surface, extending two to three times beyond these interfaces. This is a direct consequence of their large atmospheric scale height and the influence of variable gravity. The expansion of $R_{\rm{base}}$ enhances the mass loss rate by increasing the effective XUV absorption radius. In contrast, high-gravity planets have their $R_{\rm{base}}$ only slightly above $R_p$, with a negligible thickness in the radiative atmosphere. The large scale height of super-puffs results in a more gradual density gradient in the atmosphere, as shown in the bottom left panel. The top right panel illustrates that wind velocity at $R_{\rm{base}}$ in super-puffs is already significant, as their lower atmosphere is susceptible to subsonic isothermal Parker wind escape. The bottom right panel shows a stark contrast in ionization fraction: while the outflow of a normal-gravity planet becomes almost fully ionized above $R_{\rm{base}}$ (red), the outflow of a super-puff remains largely neutral across most of the wind region, despite experiencing a higher EUV flux. 

We find that for most super-puffs, $R_{\rm{base}}$ typically forms at pressure levels around 1 nbar. However, the absorption pressure decreases with lower planetary gravity and bulk density. In the lowest-gravity super-puffs modeled, $R_{\rm{base}}$ is located at 0.1 nbar, whereas for more massive Neptune-like planets, it shifts to 10 nbar. The position of $R_{\rm{base}}$ is determined by where the optical depth to XUV photons reaches unity. The optical depth at a given frequency $\nu$ is defined as:
\begin{equation}
\label{tau}
\tau_{\nu} = \sigma_\nu \int_r^{\infty} n_0dr \approx H\sigma_\nu n_0 \bigg|_{\rm{base}}
\end{equation}
where $H$ is the scale height, approximated by $kT/\mu g$; $\sigma_\nu$ is the photon cross-section at frequency $\nu$, given by $6\times10^{-18}/(h\nu/13.6\rm{eV})^3 \rm{cm^2}$; and $n_0$, the neutral number density, is approximately equal to the total number density $n$ due to the low ionization fraction in the outflow. An approximate form is given after the second equality \citep{RMC09}. Applying the ideal gas law, we find that the ionization base pressure is independent of wind temperature $T$ and can be expressed as:
\begin{equation}
\label{pbase}
P_{\rm{base}} = \frac{\mu g}{\sigma_\nu}
\end{equation}
where $\mu$ is the mean atomic weight and $g$ is the local gravity at the base. This scaling relation suggests that $P_{\rm{base}}$ is roughly proportional to gravity. However, in low-gravity planets, the precise radius calculation is highly sensitive to $P_{\rm{base}}$, underscoring the necessity of a self-consistent photoevaporation model like ours to accurately determine the wind base location.

An analytical fit (presented in Section \ref{subsec:scaling}) of our numerical results, assuming a fixed solar spectrum (see Section \ref{subsec:photo}) and applied to Eq. \ref{pbase}, suggests that the location of the EUV ionization base is primarily set by $60-70\, \mathrm{eV}$ photons for less irradiated planets ($F_{\rm{bol}}<100F_\oplus$) at ages after $\sim$10 Myr. At earlier times, this characteristic energy temporarily increases to $\sim 100\, \mathrm{eV}$. It is important to note that our results indicate super-puffs can only form within this flux range (see Section \ref{res:implication}); therefore, a consistent $60-70\,\mathrm{eV}$ photon energy sets the wind base in Eq. \ref{pbase} in this case. This energy level increases with flux in highly XUV-irradiated planets (with $F_{\rm{bol}}\ge100F_\oplus$), up to $160\,\mathrm{eV}$, with higher values observed at young ages.

In less irradiated sub-Neptunes ($F_{\rm{bol}}<100F_\oplus$), including super-puffs, EUV photons (with energy $\leq70\,\mathrm{eV}$) are the dominant driver of mass loss, while X-ray-driven escape is generally negligible. However, X-rays may contribute to mass loss in more highly irradiated sub-Neptunes, which typically have higher gravities due to their increased vulnerability to atmospheric erosion. \citet{Owen12} showed that the wind temperature achievable by X-ray heating ($T_X$) depends on the ionization parameter $\xi_X$, defined as:
\begin{equation}
\xi_{X} = 4\pi \frac{F_X}{n_{X}}
\end{equation}
where $F_X$ is the incident X-ray flux and $n_X$ is the number density at the X-ray ionization base, which is located at higher pressure levels than the EUV ionization base due to the smaller cross-section of X-ray photons. Using Equation \ref{tau}, we estimate that $\xi_X$ in super-puffs is substantially lower than $10^{-6}$, corresponding to $T_X$  well below 100 K, significantly lower than the lowest skin temperature explored in our models. This suggests that, below the EUV ionization base, the wind is predominantly heated by bolometric radiation rather than X-ray heating. We note, however, that at high atmospheric metallicity, X-ray heating may become more important due to enhanced absorption cross-sections by metal species. This possibility warrants further investigation in future work.

We then examine the energy budget. Due to the large scale height of super-puffs, we find that ionization heating is deposited almost uniformly throughout the entire subsonic outflow, resulting in a low energy input density and, consequently, a low ionization fraction. This suggests that the optical depth to EUV radiation at the sonic point is of order unity and that heating at the sonic point is significant. As a result, the wind temperature increases monotonically until reaching a layer slightly below $R_s$  (red in the top panel of Figure \ref{structure}). 

Interestingly, a similar behavior was reported for CoRoT-24c \citep{Lammer16}, but at a significantly higher gravity ($\sim 1000\ \mathrm{cm\ s^{-2}}$). Our reproduced results indicate that this apparent similarity arises from the choice of x-axis scaling in their figure, which only extends to $r/R_{\mu\mathrm{bar}} = 2$ and does not include the sonic point. If the same scale were used as in our plots, the behavior would appear consistent with the situation described below for the high-gravity case. In contrast, in high-gravity planets, ionization heating is concentrated in a negligibly thin layer at $R_{\rm{base}}$, causing a sharp rise in wind temperature. Beyond this region, $PdV$ cooling dominates, leading to a quasi-adiabatic temperature decrease with altitude toward $R_s$ (black). This effect is clearly illustrated in Figure \ref{energy}, where we compare the energy budgets of the hot Jupiter (top) and the super-puff (bottom), the same planets as in Figure \ref{structure}. Ionization heating and $PdV$ work are represented by the red solid and black dashed lines, respectively.

We find that $PdV$ work and advective cooling are the dominant cooling mechanisms for photoevaporation in low-gravity planets. Ly-$\alpha$ cooling is generally inefficient due to the low wind temperature, unless an extremely high EUV flux ($\gtrsim 10,000\ \rm{erg/s/cm^2}$) is present. However, such high fluxes are unlikely in super-puffs, as their formation requires relatively low EUV exposure (see Section \ref{res:implication}). Unlike hot Jupiters, where Ly-$\alpha$ cooling dominates at the wind base, in super-puffs, it becomes more significant at the sonic point, where temperatures are sufficiently high to trigger Ly-$\alpha$ cooling.

The temperature threshold for Ly-$\alpha$ cooling to dominate can be derived from the balance between ionization heating and Ly-$\alpha$ cooling:
\begin{equation}
F_{\rm{EUV}} \sigma_{\rm{EUV}} n_0 = An_+ n_0 \exp{(-T_0/T)}
\end{equation}
where $\sigma_{\rm{EUV}}$ is the characteristic cross-section for the EUV flux $F_{\rm{EUV}}$, $n_+$ is the ion number density, and $A=7.5\times 10^{-19}$ and $T_0=118348$ K are coefficients for Ly-$\alpha$ cooling. The exponential dependence on the right-hand side of the equation indicates that Ly-$\alpha$ cooling is only efficient at temperatures above $12,000-15,000$ K and will regulate the Ly-$\alpha$ cooling-dominated region to approximately this temperature. This threshold is slightly lower in hot-Jupiters, due to the higher $n_+$. Other cooling mechanisms, including conductive and recombination cooling, are found to be inefficient in both high- and low-gravity planets.

\subsubsection{A Comprehensive Analytical Mass-Loss model} \label{subsec:scaling}
Based on our numerical results, we present a new understanding of the scaling laws for photoevaporation in sub-Neptune mass planets. 

The energy-limited approach assumes that the mass loss rate is primarily determined by the energy available from EUV heating, given by $F_{\rm{EUV}}R_{\rm{base}}^2$ , and the energy required to overcome $PdV$ work, which has conventionally been approximated as the gravitational energy needed to lift H/He material, expressed as $GM_p\dot{M}/R_p$. This leads to the commonly used analytical mass loss rate:
\begin{equation}
\label{energy-limited}
\dot{M}_{\rm{e-lim}}  = \frac{\eta F_{\rm{EUV}}R_{\rm{base}}^2R_p}{GM_p}\approx \frac{\eta F_{\rm{EUV}}R_{\rm{base}}^3}{GM_p}
\end{equation}
where $\eta$ accounts for additional cooling sources and tidal effects. In practice, hot-Jupiter evolution models often set $R_{\rm base} = R_p$ because of the small atmospheric scale height, an approximation later adopted in many sub-Neptune studies.

However, in a full hydrodynamic framework, the $PdV$ work term can originate from multiple sources beyond just gravitational energy.  In this context, the input energy may be diverted into various sinks, including thermal energy, rather than being fully used to lift gas out of the potential well. Thus, the energy-limited escape formulation provides only an upper limit on the mass-loss rate, with the effects of additional energy sinks typically encapsulated in the efficiency factor $\eta$. Hydrostatic studies have attempted to assess the dependence of $\eta$ on planetary parameters \citep{Salz16,Owen13}, but their mass loss rate predictions fundamentally rely on the scaling of Eq. \ref{energy-limited}. Consequently, they fail to provide accurate predictions in evolutionary models with time-dependent planetary parameters or when other energy sinks dominate over gravitational dissipation. The influence of thermal energy on mass loss was qualitatively noted but not quantified in the hydrodynamic work of \citet{Erkaev16}. Here, we combine analytical derivations with numerical calculations to quantify this effect in sub-Neptunes.

The $PdV$ work can be decomposed into two terms: $d(PV)$ and $-VdP$. The first term characterizes changes in the thermal state, given by $kT/\mu$, while the second term is directly related to Euler’s momentum equation by the pressure gradient:
\begin{equation}
\label{momentum}
\rho v \frac{\partial v}{\partial r} = - \frac{\partial P}{\partial r} - \frac{GM_P \rho}{r^2}
\end{equation}
For simplicity, we ignore stellar tidal forces, as we find that photoevaporative mass loss in super-puffs is not significantly affected by them. Integrating Eq. \ref{momentum}, we find that the $-VdP$ term includes both the energy dissipated into kinetic energy $v^2/2$ and gravitational energy $GM_p/r$. Additionally, wind advection contributes to heating and cooling effects depending on the temperature gradient. This effect is non-negligible in low-gravity planets, as shown in Euler’s energy equation:
\begin{equation}
\label{euler-energy}
\rho v \frac{\partial}{\partial r}\left(c_v T\right) = \frac{kTv}{\mu} \frac{\partial \rho}{\partial r} + \Gamma
\end{equation}
where we neglect other cooling sources, including Lyman-$\alpha$ cooling, due to their minimal effects. Here, $c_v=k/(\gamma-1)/\mu$ is the specific heat capacity at constant volume, with $\gamma=5/3$ as the adiabatic index for atomic flows. The left-hand side represents advective cooling/heating, while the right-hand side consists of the $PdV$ work term and ionization heating $\Gamma = \int\limits_\nu \Gamma_\nu$, where the heating in a certain wavelength is given by:
\begin{equation}
\label{ionization-heating}
\Gamma_\nu = \epsilon F_{\nu} e^{-\tau_\nu} \sigma_\nu n_0
\end{equation}
where $\tau_\nu$ follows from the first equality in Eq. \ref{tau}, and $\epsilon = (h\nu-13.6\rm{eV})/h\nu$ represents the fraction of EUV energy converted into heat. The ideal gas law is used to link $P$, $\rho$ and $T$ , while Eqs. \ref{momentum} and \ref{euler-energy} are connected to the mass loss rate $\dot{M}=4\pi r^2\rho v$ through the mass conservation equation:
\begin{equation}
\label{continuity}
\frac{\partial}{\partial r}\left(r^2 \rho v\right) = 0
\end{equation}

\begin{figure}
\centering
\includegraphics[width=0.5\textwidth]{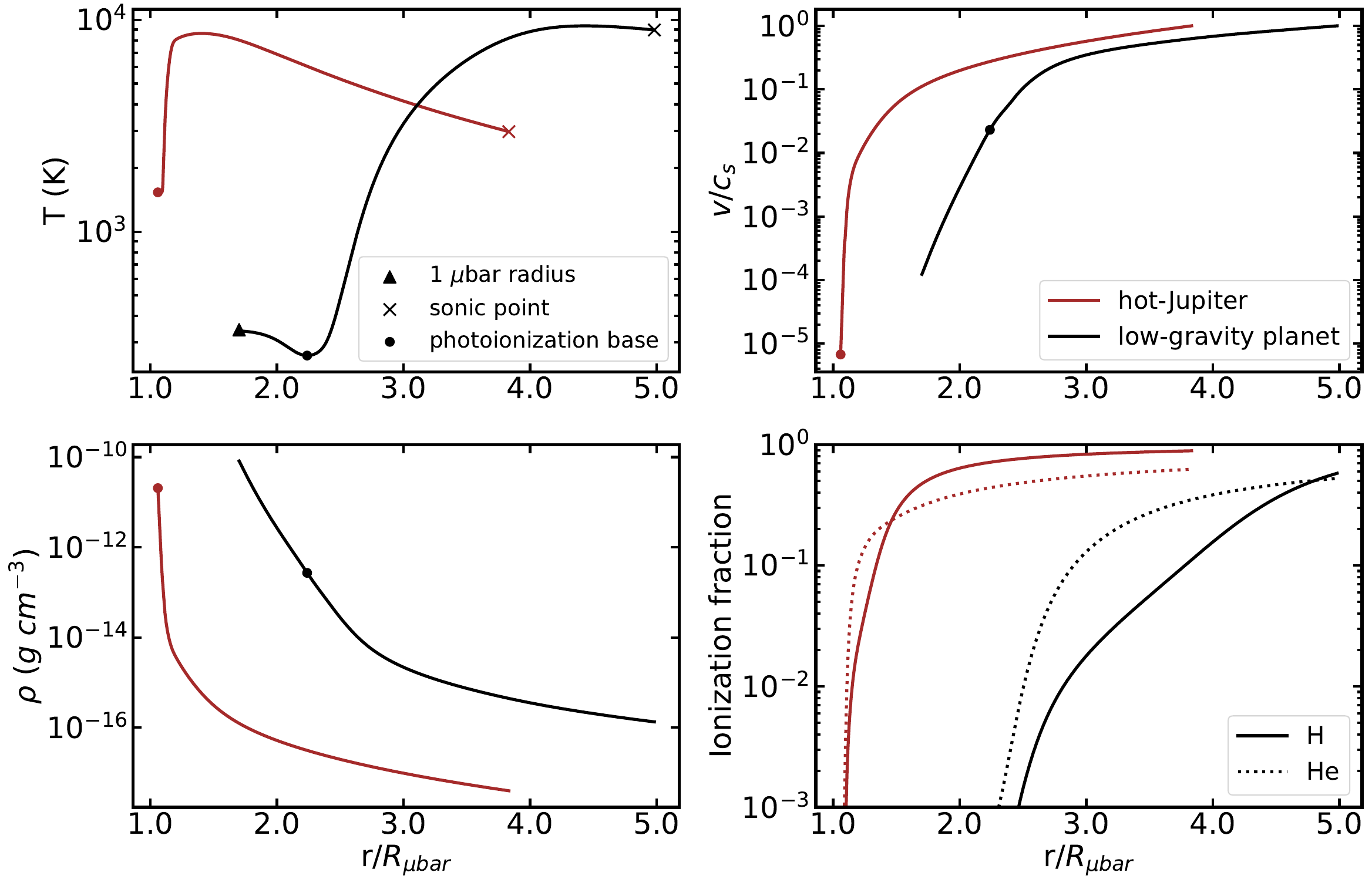} 
\caption{ Comparison of radial profiles for temperature (top left), Mach number (top right), density (bottom left), and ionization fraction (bottom right) between a super-puff (black, Kepler-51b) and a hot Jupiter (red, HD 209458b). The photoionization base is marked with circles. Structures are shown from the 1 $\mu$bar level (triangles) to the sonic point (crosses). See the main text for further discussion.
}
\label{structure}
\end{figure}

\begin{figure}
\centering
\includegraphics[width=0.5\textwidth]{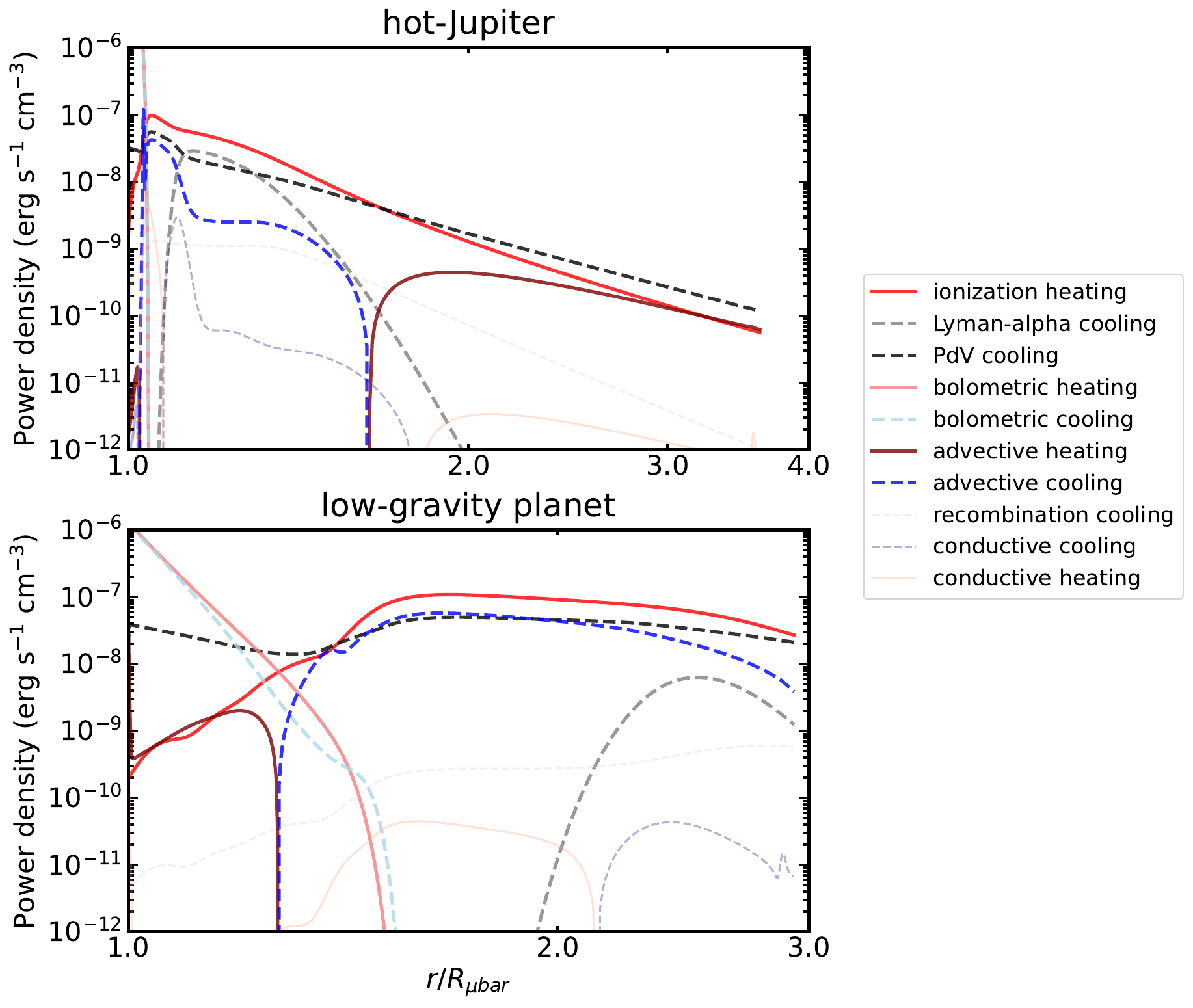} 
\caption{ Comparison of energy budgets for high-gravity (top) and low-gravity (bottom) planets, corresponding to the same planets shown in Figure \ref{structure}. Refer to the main text for a detailed discussion.
}
\label{energy}
\end{figure}

Integrating Eqs. \ref{momentum}, \ref{euler-energy}, and \ref{continuity} yields the energy budget of the subsonic wind below $R_s$ in Bernoulli’s form:
\begin{equation}
\label{bernoulli}
\begin{split}
\pi\int\limits_\nu\int_{R_{\rm{base}}}^{R_s} \Gamma_\nu r^2 dr d\nu &= \dot{M} \big[(\frac{GM_p}{R_{\rm{base}}}-\frac{GM_p}{R_s}) + (\frac{k T_s}{\mu_s} - \frac{k T_{\rm{base}}}{\mu_{\rm{base}}}) + \\
&(c_{v,s}T_s - c_{v,\rm{base}}T_{\rm{base}}) + (\frac{1}{2}c_s^2-\frac{1}{2}v_{\rm{base}}^2)\big] \\
&\approx \dot{M} \big[\frac{GM_p}{R_{\rm{base}}} + c_{p,s}T_s + \frac{1}{2}c_s^2\big] \\
&= \dot{M} \big[\frac{GM_p}{R_{\rm{base}}} + (\frac{\gamma}{\gamma-1}+\frac{\gamma}{2}) \frac{kT_s}{\mu_s}\big]
\end{split}
\end{equation}
where $c_s=\sqrt{\gamma k T_s/\mu_s}$ is the sound speed at the sonic point. This equation includes all four contributions discussed earlier. Apart from the gravitational energy dissipation term, which corresponds to the squared escape velocity, all other terms scale with the thermal energy $k T_s/\mu_s$, where $T_s$ and $\mu_s$ are the temperature and mean molecular weight at the sonic point. The second equality follows from the facts that $R_s$ is a few times larger than $R_{\rm{base}}$, the wind base temperature $T_{\rm{base}}$ (equal to the skin temperature $\sim T_{\rm{eq}}$) is much smaller than $T_s$, and the velocity at the base $v_{\rm{base}}$ is negligible compared to $c_s$ (see Figure \ref{structure}). For convenience, we combine the advective cooling/heating term and the thermal state change term into a single expression for the change in specific enthalpy, where the specific heat capacity at constant pressure is given by $c_p = \gamma/(\gamma-1)*k/\mu$.

The photoionization heating term on the left-hand side of Eq. \ref{bernoulli} cannot be integrated analytically, as isothermal conditions are not applicable to the outflow of super-puffs. Our numerical results suggest that, to first order, this term can be approximated as $\alpha \pi F_{\rm{EUV}} R_{\rm{base}}^2$, where the order-unity coefficient $\alpha$ is determined to be 0.45. This coefficient exhibits only a marginal dependence on gravity and EUV flux. We thus demonstrate that traditional energy-limited escape occurs when thermal energy terms are negligible, a condition expected for high-gravity planets. 
We note that the efficiency factor $\eta$ remains necessary to account for the tidal forces and additional cooling mechanisms, including Lyman-$\alpha$ cooling. We determine $\eta$ in Eq. \ref{energy-limited} to be in the range $0.3-0.5$ for low-gravity sub-Neptunes, which is higher compared to the commonly adopted range of $0.1-0.3$. A fixed 0.1 is used in the evolution models of \citet{Lopez12,Howe15,Chen16,Owen17}, whereas a variable $\eta$ is discussed in \citet{Owen13}. This higher efficiency arises because low-gravity sub-Neptunes are less susceptible to spectrum line cooling.

For low-gravity planets, the dominant mechanism is within a new mass-loss regime, which we term ``thermal-energy-mediated'' photoevaporation (TEMP): the EUV energy input is primarily converted into kinetic energy via $PdV$ work, and used to sustain the wind temperature (which drives the pressure gradient) through advective energy transport. Note that while these thermal energy terms do not directly alter the planetary radius (i.e., through ``lifting''), they are essential for photoevaporation: the gas must be sufficiently heated and accelerated to the sound speed in order to overcome the gravitational potential well. In this sense, the outflow is still limited by the input energy, but critically set by different physical formulations. Accordingly, we propose the following scaling law:
\begin{equation}
\label{mdot-scaling}
\dot{M} = 
\begin{dcases}
    \frac{\eta F_{\rm{EUV}}R_{\rm{base}}^3}{GM_p},     & \text{if } \lambda_s \geq 1\\
    \frac{\pi \alpha F_{\rm{EUV}} R_{\rm{base}}^2 \mu_s}{(\frac{\gamma}{\gamma-1}+\frac{\gamma}{2})kT_s} ,     & \text{if } \lambda_s < 1\\
\end{dcases}
\end{equation}
where the atmospheric escape parameter at the sonic point, $\lambda_s \equiv (GM_p/R_s)/(kT_s/\mu_s)$, quantifies the relative significance of gravitational and thermal energy and sets the transition between the two regimes. The second expression follows from rearranging Eq. \ref{bernoulli}, neglecting the gravitational potential energy term.

Since determining the planet's mass-loss regime relies on knowledge of the escape parameter $\lambda_s$ itself, we provide an empirical approach for estimating $\lambda_s$. First, we find that for any PdV-dominated EUV-driven wind, the temperature $T_s$ follows a scaling relation with the EUV flux:
\begin{equation}
\label{temperature-scaling}
T_s = \beta F_{\rm{EUV}}^{0.4}
\end{equation}
Fitting our numerical results, we find that the coefficient $\beta$ depends weakly on gravity and can be approximated as a constant, $\beta \approx 200$. See Appendix \ref{app:fit} for details of the fitting.

Additionally, we find that in the outflows of low-mass planets, ionization balance is dominated by ion advection, while recombination remains negligible due to the low electron number density. This leads to the ionization balance equation:
\begin{equation}
\label{ionization}
\int\limits_\nu \frac{F_{\nu} e^{-\tau_\nu}}{h\nu} \sigma_\nu n_0 d\nu = nv \frac{\partial f_+}{\partial r}
\end{equation}
where $n$ is the total number density and $f_+$ represents the ionization fraction for a single species. Integrating Eqs. \ref{momentum}, \ref{euler-energy} and \ref{ionization} while neglecting gravitational dissipation, we derive the ionization fraction scaling in the thermal-energy-mediated photoevaporation regime:
\begin{equation}
\label{ionization-scaling}
f_{+,s} = T_s/\phi
\end{equation}
\begin{equation}
\phi = \frac{h}{{k\left(\frac{\gamma}{\gamma-1}+\frac{\gamma}{2}\right)}}\frac{\iint\limits_{\nu,\tau_\nu} \epsilon F_\nu e^{-\tau_\nu}d\tau_\nu d\nu}{\iint\limits_{\nu,\tau_\nu} \frac{1}{\nu} F_\nu e^{-\tau_\nu}d\tau_\nu d\nu}
\end{equation}
where the coefficient $\phi$ is on the order of $h\nu/k\sim 11,000$, with order-unity factors that depend on the atomic species and the XUV spectrum. We find $\phi$ to be insensitive to gravity and flux. Numerically, we determine $\phi$ to be 15,000 for hydrogen and 18,000 for helium. Consequently, the mean molecular weight at the sonic point, $\mu_s$, is given by:
\begin{equation} 
\label{mmw}
\mu_s = \frac{m_H}{(1-Y)(1+f_{+,s,H})+Y(1+2 f_{+,s,He})/4} 
\end{equation}
where $Y$ denotes the helium mass fraction.

Incorporating Eq. \ref{temperature-scaling}, the thermal-energy-mediated mass loss rate in Eq. \ref{mdot-scaling} can be expressed analytically as:
\begin{equation}
\label{thermal-limited-mdot}
\dot{M} = 6.6\times 10^{8} g\ s^{-1}\left(\frac{F_{\rm{EUV}}}{1000\ \rm{erg\ s^{-1} \ cm^{-2}}}\right)^{0.6}\left(\frac{R_{\rm{base}}}{R_\oplus}\right)^2 \frac{\mu_s(F_{\rm{EUV}})}{m_H}
\end{equation}
where we assume $\gamma=5/3$ for an atomic H/He mixture, and $\mu_s$ is determined via Eq. \ref{temperature-scaling}, \ref{ionization-scaling} and \ref{mmw}. This scaling is consistent with the analytical fit derived from 3D hydrodynamic simulations in \citet{Wang19}. Eq. \ref{thermal-limited-mdot} can be readily implemented in other sub-Neptune evolution models to improve mass loss estimates.

To estimate $\lambda_s$ at the transition, we analytically fit our model results and find that the $R_s/R_{\rm{base}}$ ratio follows a quadratic function, $y=a_2 x^2 + a_1 x + a_0$, of $\log(F_{\rm{bol}}/F_\oplus)$, independent of gravity and stellar parameters, where $R_{\rm{base}}$ can be analytically determined using Eq. \ref{pbase}. The best-fit coefficients are $a_2=-0.147$, $a_1=-0.315$ and $a_0=3.914$. This parameterization applies exclusively to boil-off-constrained planets (rather than those insusceptible to boil-off) and is valid for F- and G-type stars. See Appendix \ref{app:fit} for other details. 

Estimating $R_{\rm{base}}$ is essential for determining the mass-loss rate via Eq. \ref{mdot-scaling}. Analytically, the corresponding $P_{\rm{base}}$ can be obtained from Eq. \ref{pbase}, where the gravity at the base should be evaluated following our treatment of the lower boundary condition in Section \ref{subsec:photo} (see Figure \ref{transition}). To determine the characteristic photon energy that defines the wind base, we provide an analytical fit calibrated to our hydrodynamic results, in the form of:
\begin{equation}
\nu_{\rm{base}}=b_1(\frac{F_{\rm{EUV}}}{g^{1.2}})^{b_2}+b_0
\approx b_1^{\prime} \frac{E_{\rm{th}}}{E_{\rm{g}}}\sqrt{\frac{M_p}{M_\oplus}}+b_0
\end{equation}
where $g$ is the gravity evaluated at the RCB. The best-fit coefficients are $b_0=55.82$, $b_1=14.48$ and $b_2=0.4088$. This fit can be rewritten into the second equality using Eq. \ref{temperature-scaling}, where $E_{\rm{th}}=kT_s/m_{H}$ and  $E_{\rm{g}}=GM_p/R_{\rm{RCB}}$ are the characteristic thermal energy and gravitational energy of the outflow, respectively. $b_1^{\prime}$ is determined to be $17.51$. This form highlights that $\nu_{\rm{base}}$ remains nearly constant across planets in the energy-limited regime with $E_{\rm{th}}/E_{\rm{g}} \ll 1$ (see Appendix \ref{app:fit}).

Together, these scaling laws fully determine the mass-loss rates for sub-Neptunes. We find that this analytical method reproduces our hydrodynamic results with high accuracy in evolutionary calculations. Nevertheless, all results presented below are based on the more comprehensive numerical model with hydrodynamics.

\begin{figure}
\centering
\includegraphics[width=0.45\textwidth]{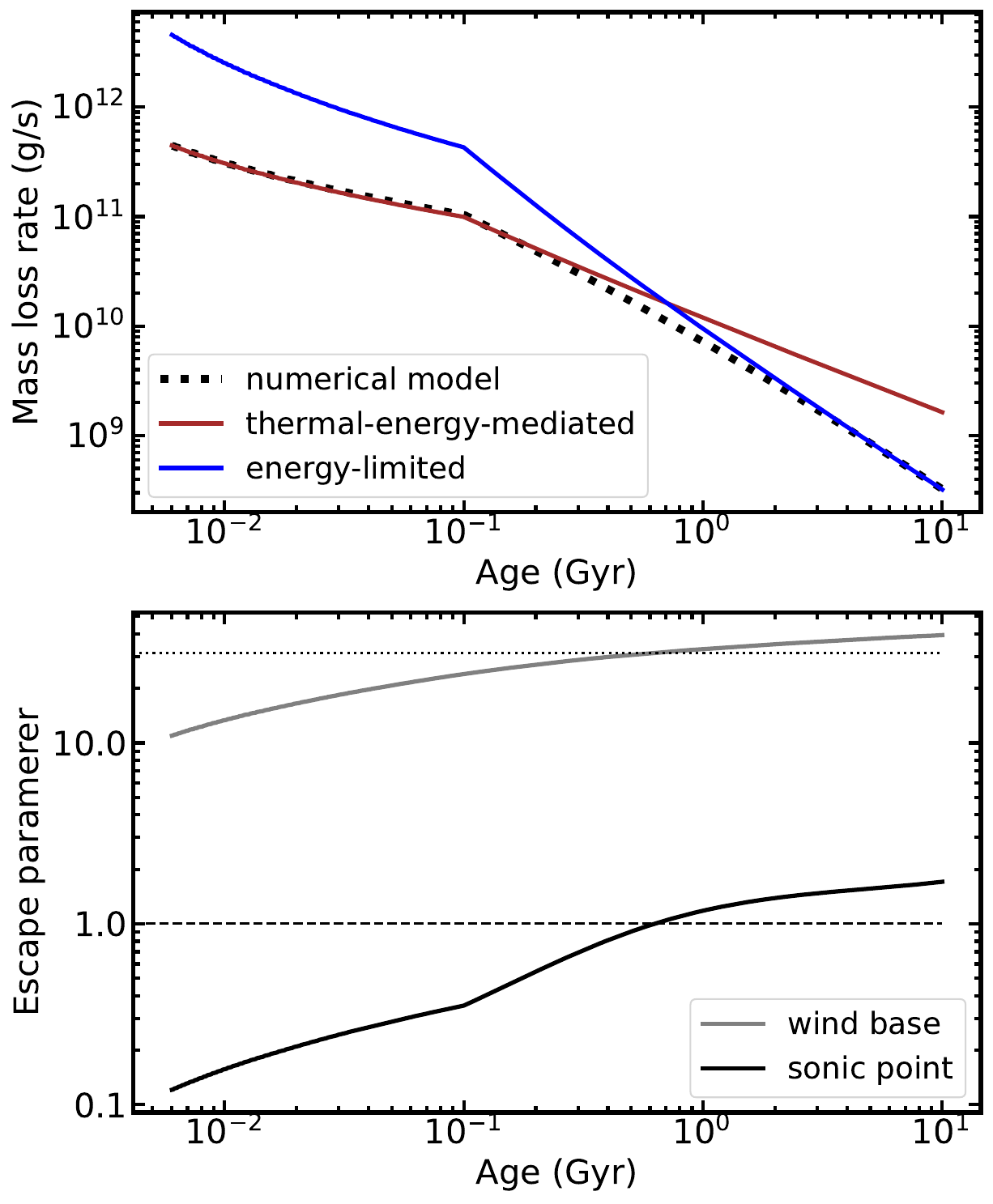} 
\caption{ In the top panel, we display the thermal-energy-mediated (red) and energy-limited (blue) mass loss rates, based on Eqs. \ref{thermal-limited-mdot} and \ref{energy-limited}, respectively. For comparison, the mass loss rate from our numerical model is shown with a black dashed line. The kinks in the mass-loss rate reflect the evolution of the EUV flux, which remains saturated during the early stages. In the bottom panel, we plot the evolution of the escape parameter at the sonic point ($R_s$, black) and at the wind base ($R_{\rm{base}}$, gray). 
}
\label{scaling}
\end{figure}

In Figure \ref{scaling}, we compare mass loss rates from our numerical model (dashed lines) with the analytical scalings for both the thermal-energy-mediated (red) and energy-limited (blue) regimes in the top panel. We confirm an excellent agreement across all models presented in this study. The bottom panel depicts the evolution of $\lambda_s$, demonstrating that the transition between the two regimes occurs precisely when $\lambda_s$ reaches unity.  This typically corresponds to a wind-base $\lambda$ value of 30 or to a factor-of-two reduction in the envelope size following boil-off. This finding implies that sub-Neptunes susceptible to boil-off always enter the thermal-energy-mediated regime before transitioning to the energy-limited phase. Additional planet samples are provided in Appendix \ref{app:fit}.

\subsubsection{Comparison to other Hydrodynamic Escape models}
\citet{RMC09} identified another mass loss regime, termed ``recombination-limited'' escape, relevant for hot Jupiters. In this regime, intense EUV irradiation heats the outflow to $\sim$10,000 K, where recombination-driven Ly-$\alpha$ cooling effectively thermostats the entire wind to be isothermal (see also \citet{Owen16}). Here, the ionization fraction is governed by recombination, which in turn sets the mass density at the photoionization base. Since a higher EUV flux increases this density, the resulting mass loss rate follows a scaling of $F_{\rm{EUV}}^{0.6}$. 

Despite this similarity in scaling with $F_{\rm{EUV}}$, we argue that recombination-limited mass loss is fundamentally distinct from thermal-energy-mediated photoevaporation. Specifically, we find that recombination-limited escape is irrelevant for super-puffs, as the isothermal sonic point at 10,000 K lies below the photoionization base. For recombination-limited escape to be physically viable, a high gravity is necessary: the planetary mass must be at least $20M_\oplus$ so that its isothermal sonic point (Eq. \ref{sonicpoint}) lies above its rock/iron core and this threshold rises to $100M_\oplus$ for a gas giant. Conversely, for thermal-energy-mediated photoevaporation to occur, we estimate an upper mass threshold of $35M_\oplus$.

Here, we propose that the energy-limited and thermal-energy-mediated regimes can be collectively termed PdV-dominated photoevaporation, in which the mass-loss rate is primarily governed by the balance between $PdV$ work and the input energy. In contrast, the recombination-limited regime is best characterized as radiative-cooling-dominated escape.

TEMP should not be confused with core-enhanced photoevaporation \citep{Owen24}, in which core luminosity maintains an inflated lower-boundary radius, coupling thermal evolution and atmospheric escape (see also \citet{Ginzburg16,Tang24}) and thereby boosting the mass-loss rate. This effect is self-consistently included in our framework through interior evolution modeling. In contrast, TEMP arises from the thermal energy conversion in the outflow, which alters the wind structure rather than the lower boundary condition.

\citet{Kubyshkina18b} presented an analytical scaling for mass loss in the low-gravity regime. The transition between their low- and high-gravity regimes (their Eq. 10) occurs at an escape parameter $\lambda \lesssim 10$, which corresponds to boil-off \citep{Fossati17,Tang25}. In this and related works \citep{Kubyshkina18b,Kubyshkina18,Krenn21}, this escape was attributed to ``planetary intrinsic thermal energy'' from contraction, although it was later shown to be stellar bolometric-driven thermal energy \citep{Misener21,Tang24,Rogers24} combined with internal energy release from an adiabatic process in the envelope \cite{Tang24}) coupled with low gravity. This results in mass-loss rates exceeding the energy-limited prediction, which can only be short-lived evolution-wise. Their (boil-off) escape regime has a quite different dependence on radius, mass and XUV flux compared to TEMP. In contrast, we find that TEMP is sustained entirely by XUV photons for an extended period of time after boil-off and yields rates well below the energy-limited value.

\subsection{Comparison to Previous Evolution Models} \label{disc:comp}
We suggest that previous evolution and population studies have significantly overestimated the role of photoevaporation in low-gravity plants ($g\lesssim200 cm\ s^{-2}$) while underestimating planetary radii, especially for low-mass and low-density planets.  This has made it challenging to understand super-puffs. Three key factors contributed to these uncertainties:

First, previous models did not explicitly account for a boil-off phase preceding photoevaporation. As a result, they overestimated planetary radii at young ages by assuming excessively high initial H/He mass fractions, which in turn led to exaggerated photoevaporation mass loss rates (see also \citet{Kubyshkina18b}). \citet{Rogers21} attempted to include a boil-off phase by initializing their models with a post-boil-off mass fraction determined from the analytic scaling of \citet{Ginzburg16}. However, \citet{Tang24} found that this scaling underestimates the dependence on planetary mass, leading to an overestimation of post-boil-off mass fractions in low-mass planets. 

Second, the widely used ``energy-limited'' formulation (Eq. \ref{energy-limited}) for photoevaporation in low-mass sub-Neptunes systematically overestimates mass loss rates at young ages. In Figure \ref{scaling}, we demonstrate that this scaling (blue) predicts mass loss rates roughly ten times higher than those derived from the more accurate thermal-energy-mediated formulation (red), given the same $R_{\rm{base}}$. This discrepancy arises from the neglect of the thermal energy term, $kT_s/\mu_s$, which is minor (10\% of the gravitational potential energy, $v^2_{\rm{esc}}$) in hot-Jupiters, but can exceed several times $v^2_{\rm{esc}}$, in low-gravity planets at young ages, as seen in the bottom panel of Figure \ref{scaling}. Several studies \citep{Jin14,Owen16,Lopez18,Owen24} have attempted to mitigate the mass-loss overestimate by incorporating recombination-limited photoevaporation at young ages. However, as shown in Section \ref{res:photo}, recombination-limited mass loss is relevant only for high-gravity planets and at extremely high EUV fluxes \citep[$>10^5\ \rm{erg\ s^{-1} \ cm^{-2}}$, ][]{RMC09}. We find that applying recombination-limited mass loss to low-density sub-Neptunes still results in mass-loss overestimation.

Moreover, these models relied on a simplified hydrostatic radiative atmosphere with constant scale heights. This oversimplification introduces substantial underestimation in radius calculations (see \citet{Tang25}), on which the mass rate sensitively depends. This leads to an underestimation of the mass loss. 

\begin{figure}
\centering
\includegraphics[width=0.45\textwidth]{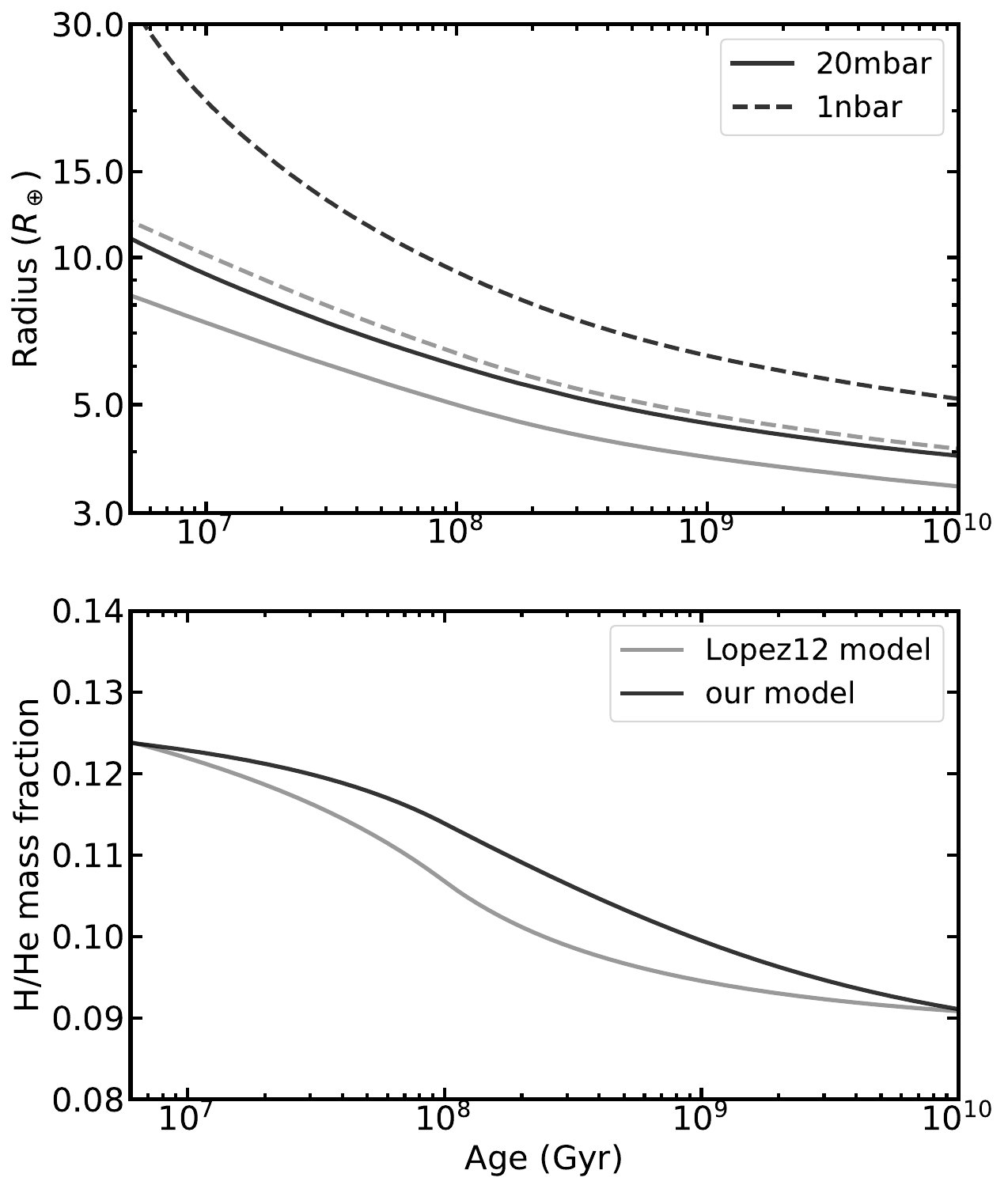} 
\caption{ Comparison of planetary evolution for Kepler-51b during the photoevaporation phase between our new framework (black) and the previous framework (gray), which assumes a constant scale height for the radiative atmosphere and energy-limited escape. The top panel presents the 20mbar (solid) and 1nbar (dashed) radii. For consistency in comparison, both models start with the same initial conditions, self-consistently calculated from a boil-off phase. In contrast, the previous framework treats initial conditions as free parameters, whose effects are not shown in the figure. See text for further discussion.
}
\label{comparison}
\end{figure}

Figure \ref{comparison} compares the evolution of planetary radii (top) and envelope mass fractions (bottom) between our framework and the model of \citet{Lopez12}, which behaves similarly to many previous models. Note that we have ignored the contribution from boil-off by setting the same initial conditions between models. While the discrepancies in the mass-loss and radiative atmosphere modeling result in similar final mass fractions (bottom) for this specific planet (Kepler-51b), the planetary radius remains underestimated by more than a factor of two in the previous models (top). While the relative difference in the final mass fraction varies between planets, we found a systematic underestimation of the planetary radii in previous models. Consequently, the larger planetary radii from our model significantly aid in explaining super-puffs, as demonstrated in the sections below.

\subsection{Application to Observed Super-Puffs} \label{res:app}
In this section, we apply our novel fully coupled numerical evolution model to observed super-puffs and young (low gravity) planets listed in Table \ref{table:para} to validate our model. For the Kepler-51 system, a recent update \citep{Masuda24} provides two possible sets of planetary masses depending on whether the newly discovered Kepler-51e lies inside or outside the 2:1 resonance with Kepler-51d. We vary parameters such as planetary mass, system age, and transit pressure level to match the observed mass, age, and radius within their respective observational uncertainties. However, accounting for uncertainties in stellar parameters, including stellar mass, radius, and luminosity (or effective temperature), is beyond the scope of this study.

\begin{figure*}
\centering
\includegraphics[width=0.9\textwidth]{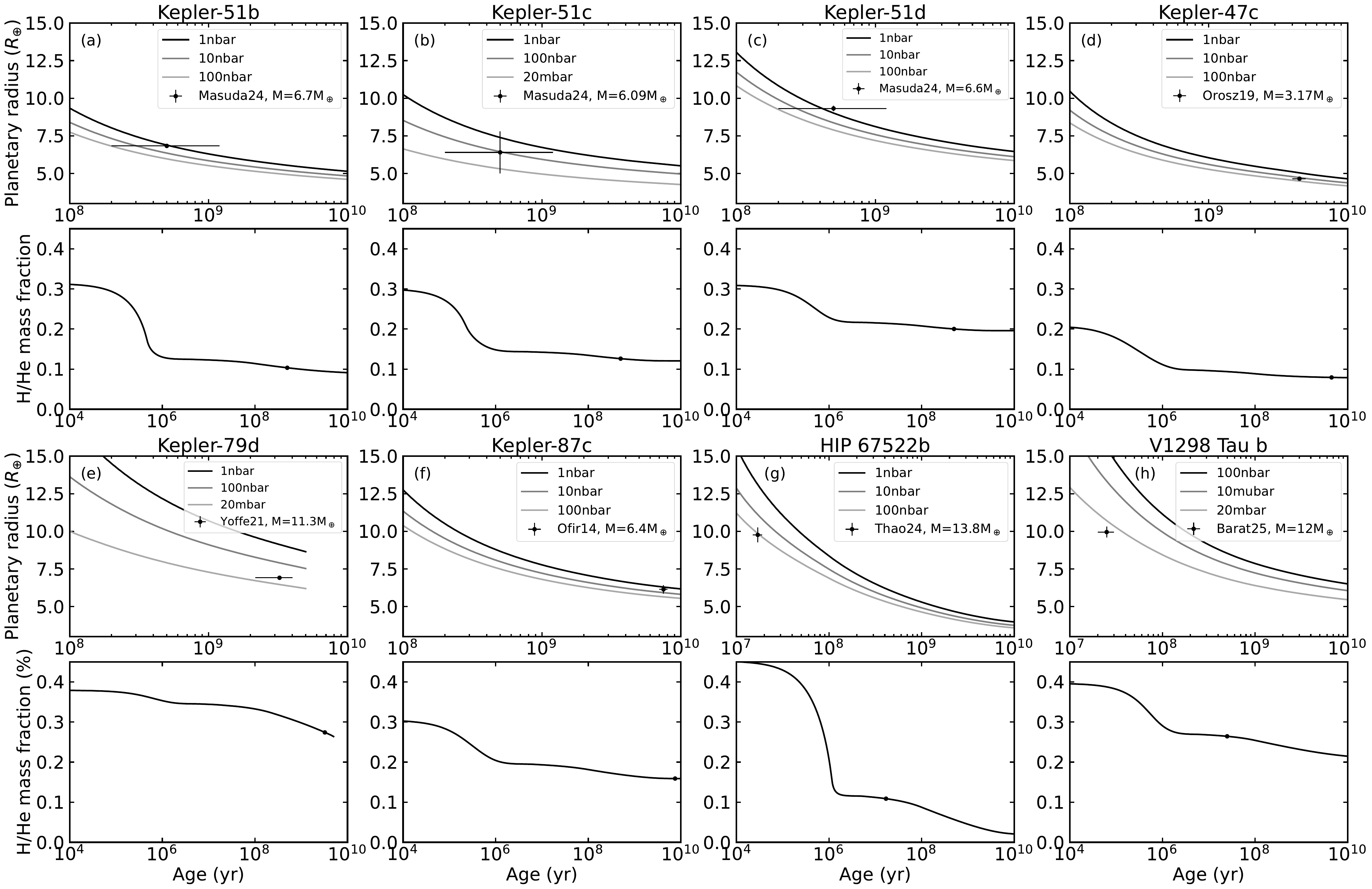} 
\caption{ Modeled evolutionary tracks of planetary radii and H/He mass fractions (defined as the ratio of H/He mass and total planetary mass) for the sample super-puffs and young planets in Table \ref{table:para}. Note that the x-axis scale varies between panels, reflecting differences in cooling and mass loss timescales. Radii at various pressure levels are shown, along with observational data points for comparison. For clear atmospheres, the transit pressure typically lies at the mbar level, while hazy atmospheres push this pressure down to the nbar range. Planetary age is defined as the time elapsed since the dispersal of the inner disk. Circles in the H/He mass fraction panels indicate the epoch corresponding to the observational data.
}
\label{tracks}
\end{figure*}

First, we have run suites of evolution models for the planets in Table \ref{table:para}, to match their observed mass and radius at the given system age. In Figure \ref{tracks}, we present these tracks, showing planetary radii at different pressure levels (first rows) and the corresponding H/He mass fractions (second rows). The best-fit values of these planets are  consistent with our model-predicted radii within the reasonable pressure ranges (1-100nbar for a hazy atmosphere and 20mbar for a clear atmosphere). Our results indicate that the Kepler-51 planets and V1298 tau b will remain puffy for the next several billion years, whereas HIP 67522b (which is quite young) will lose its super-puff characteristics on Gyr timescales due to vigorous photoevaporation.

\begin{figure*}
\centering
\includegraphics[width=0.9\textwidth]{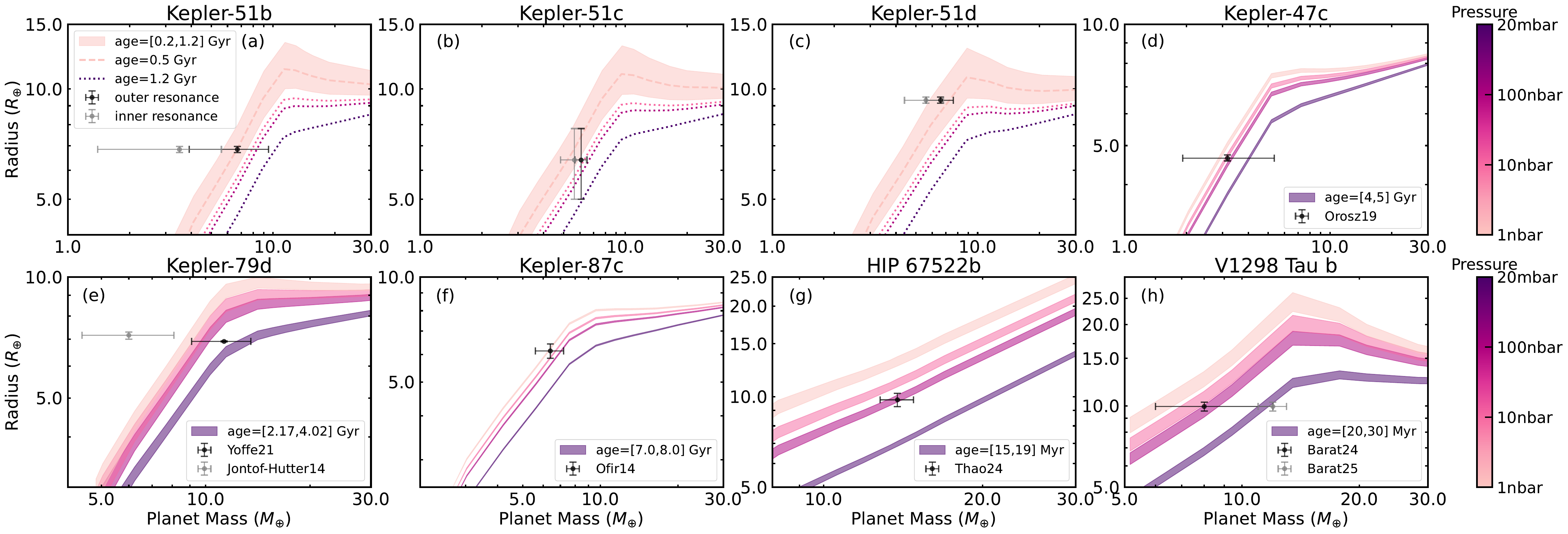} 
\caption{ M-R curves for the selected super-puffs and young planets. We show planetary radii at different pressure levels (color-coded). The width of each shaded region reflects the variation in radius due to uncertainties in system age. For Kepler-51 planets, only the lower-bound radii at pressure levels other than 1 nbar are shown (dotted). We also display the 1 nbar radius at the best-fit system age (dashed). See the main text for further discussion.
}
\label{mrk}
\end{figure*}

To account for uncertainties in planetary mass and further validate our model, we vary rock/iron core mass and track planetary radii within the age range constrained by observations. Figures \ref{mrk} present these mass-radius (M-R) relationships at various pressure levels (20mbar-1nbar, color-coded) for the stellar and orbital parameters of the observed super-puffs and young planets. Note that the x-axis denotes the total planetary mass rather than the rock/iron core mass. The width of each shaded region reflects radius variations due to age uncertainty. For comparison, we overlay the observed masses and transit radii along with their uncertainties, allowing us to constrain possible transit pressure levels in the context of our model.  Panels d) through h) show this most clearly.  In panels a), b), and c), these shaded regions for different pressure levels would significantly overlap. To avoid overlap for these Kepler-51 planets we only show the full shaded region at 1 nbar.  For the other transit pressures, we show the lower-bound radius determined by the upper-bound system age, denoted by dotted curves. For these planets, we additionally show the 1 nbar radius at the best-fit system age of 0.5 Gyr in dashed curves.

A transit pressure of less than 1 nbar is considered unlikely. Given the role of outflows in lifting haze particles to high altitudes \citep{Wang19}, the altitude corresponding to this pressure level is typically above the photoionization base. This would inhibit photoevaporation, contradicting the underlying assumptions of our model.

Since our model does not track envelope-free evolution of bare rock-iron planets, we truncate the M-R curves at low masses, where planets lose their entire convective H/He envelopes. \citet{Tang25} found that by the time a planet becomes envelope-free, its remaining H/He mass fraction is typically on the order of 0.01\%. The precise value depends on the incident bolometric flux: highly irradiated planets tend to have deeper RCBs, leading to envelope loss at a higher H/He mass fraction. Additionally, higher flux levels shift the truncation to higher planetary masses. 

We find excellent agreement between our model and observations, favoring certain TTV mass measurements over others, as discussed below. For Kepler-51b, we find that the mass determined under the assumption of ``inner resonance'' (gray sample points) in \citet{Masuda24} is too low to account for its large radius. Instead, we suggest that the masses determined under the ``outer resonance'' assumption (black sample points) provide a great match to our model: Kepler-51d falls well within the 1 nbar radius range, suggesting a hazy atmosphere, which is consistent with new spectral observations \citep{Roberts25}. On the other hand, for both Kepler-51b and Kepler-51c, due to large observational uncertainties, we cannot precisely constrain their transit pressures, although they appear to align better with nbar-level pressures given the current mass measurements. However, recent \emph{JWST} transmission spectroscopy reveals clear atmospheric absorption features (Peter Gao, personal communication), suggesting that Kepler-51b likely has a higher mass, estimated to be 9–10$M_\oplus$ to match the 20 mbar radius (although the transit pressure could be lower). 

For Kepler-87c, based on the existing mass measurements, our model strongly suggests a hazy atmosphere. We suggest follow-up spectroscopic observations to confirm this prediction. For V1298 Tau b, \citet{Barat24} provide evidence of a clear atmosphere, which, based on our model, suggests a higher planetary mass than the initial measurement, to match the 20 mbar curve in the bottom right panel. A revised mass estimate from Barat et al. (2025, in review) of $12_{-1}^{+1} M_\oplus$ closely matches our prediction. 

We find that all other planets best fit the nbar pressure levels, suggesting they likely have hazy atmospheres, except for Kepler-79d (panel e), which is more likely to have a clear atmosphere. Due to large observational uncertainties, we cannot rule out the possibility of the opposite atmospheric classifications for Kepler-47c, Kepler-79d, and V1298 Tau b. Among the uncertainties in age, mass, and radius, we identify mass measurement as the dominant source of uncertainty, as it critically determines the mass loss rate and, consequently, the planetary radius. 

The alignment of the radii of these young planets and super-puffs suggests that they were likely born with a substantial reservoir of H/He gas in their envelopes and underwent a boil-off phase. This agreement also provides strong validation for our model. Our results further indicate that super-puffs can accrete and retain sufficient H/He mass at their current orbital positions, implying that the inward migration scenario proposed by \citet{Lee16} may not be required.

\begin{figure}
\centering
\includegraphics[width=0.5\textwidth]{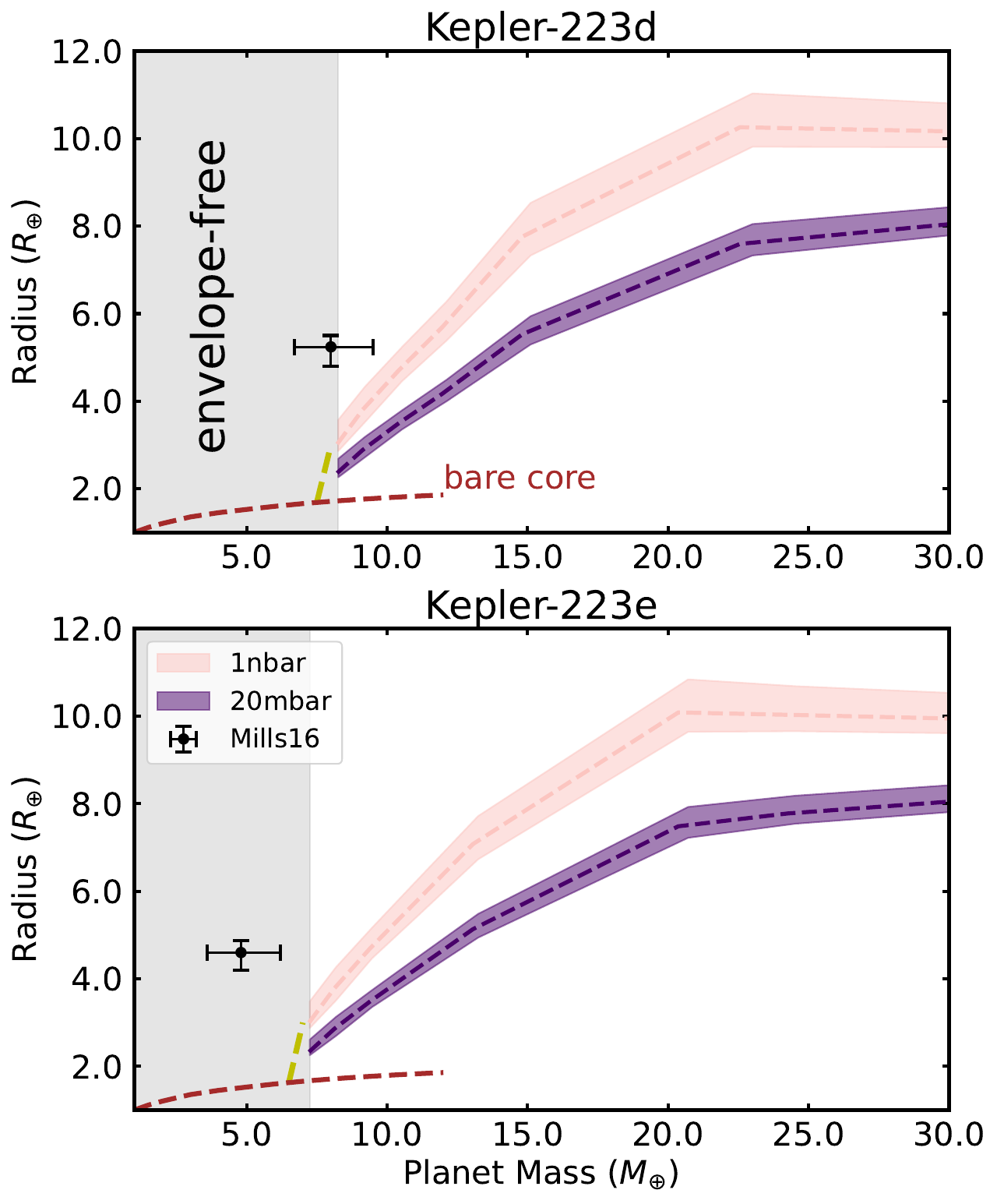} 
\caption{ M-R curves for Kepler-223 planets. The 1 nbar (pink) and 20 mbar (purple) radii correspond to hazy and clear atmospheres, respectively. The envelope-free phase is indicated by the gray shaded region. Planets are expected to evolve rapidly toward a thin-atmosphere rocky state, represented by the red dashed curve, following the green dashed trajectory. Observational uncertainties in mass and radius are shown as error bars on the black data points, while uncertainties in age contribute to the dispersion of the M-R curves. The best-fit age is marked by dashed lines. Our model results suggest that the currently reported mass measurements for the Kepler-223 planets are inconsistent with their expected evolutionary paths and likely underestimate the true masses.
}
\label{mr223}
\end{figure}

We also identified two super-puffs inconsistent with our model. In Figure \ref{mr223}, we show that the current mass measurements place both Kepler-223d and Kepler-223e in the envelope-free phase (marked with gray shades), making it impossible for them to retain enough envelope mass to reach their observed radii. Based on the observed radii and system age, we estimate that these planets must have minimum masses of at least 10$M_\oplus$ and 8$M_\oplus$ if they possess a hazy atmosphere, and 13$M_\oplus$ and 10$M_\oplus$ if they have a clear atmosphere, where the 1 nbar radius and 20 mbar radius are shown in pink and purple, respectively. Therefore, we suggest a new TTV analysis with improved physical assumptions (if possible) to reassess their masses in future work.

\citet{Tang25} suggests that after losing the H/He envelope, a low-mass super-Earth can retain a residual thick radiative H/He atmosphere due to its large atmospheric scale height \citep{Tang25}. In Figure \ref{mr223}, we compare this with the bare rock/iron core radius (red), which are 30–50\% smaller than the transit radius at the transformation boundary, the largest radius an envelope-free planet can attain, depending on the pressure level. This boundary effectively marks the upper edge of the radius valley. The transition from an envelope-free state to a bare rock/iron core with an Earth-like thin atmosphere likely occurs within a narrow mass range: \citet{Tang25} demonstrated that the final mass fraction after boil-off decreases sharply with decreasing core mass, and such small residual H/He masses are readily lost to photoevaporation during subsequent evolution. An illustrative M-R curve representing this transition regime, plotted in olive green in both panels, is not computed from models but rather a schematic extrapolation using a steeper slope.

\begin{figure*}
\centering
\includegraphics[width=0.9\textwidth]{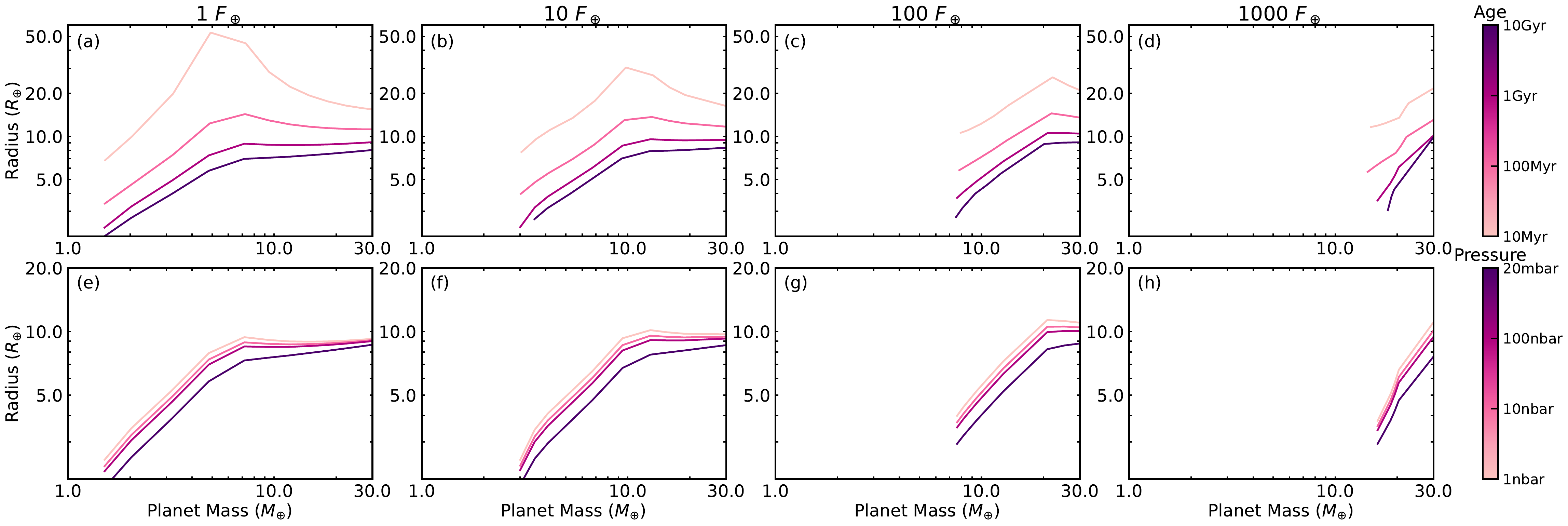} 
\caption{ M-R curves for planets orbiting Sun-like stars, shown as a function of incident flux (different columns), age (top row), and pressure level (bottom row). The pressure level is fixed at 10 nbar in the top panels, and the age is fixed at 500 Myr in the bottom panels. Notably, we find that boil-off in planets receiving 1000$F_\oplus$ is highly sensitive to mass, resulting in distinct kinks in the curves. This behavior warrants further physical investigation. The transformation boundary, marking the transition to an envelope-free state, results in the truncation on the left. Accurately determining this boundary requires a large set of models with varying masses and ages; thus, we have only calculated this boundary with a precision of 0.5$M_\oplus$.
}
\label{mrs}
\end{figure*}

Lastly, Figure \ref{mrs} generalizes the M-R curves from Figure \ref{mrk} by varying incident flux (different columns) and system age (top row, color-coded), both of which strongly influence the time-integrated mass loss and, consequently, planetary radii. Since most super-puffs orbit around Sun-like stars, this analysis focuses specifically on such hosts. Other details are available in the figure caption.

For all modeled M-R curves, the planetary radius increases linearly with mass in log-log space. This positive slope arises because a heavier core retains a more massive envelope against boil-off and photoevaporation. The curve transitions to a flatter section once a planet exceeds a mass threshold where it is no longer sensitive to boil-off. As shown in Figure \ref{mrk} and \ref{mrs}, young planets exhibit a negative slope due to the increasing variable gravity effect with decreasing mass, whereas older planets display a positive slope since they retain more H/He from formation and are less susceptible to boil-off and photoevaporation. Notably, for HIP 67522b, planets remain sensitive to boil-off even beyond 30$M_\oplus$ due to their high irradiation levels.

\subsection{Implications for Super-Puff Observations}
\label{res:implication}
The population study in \citet{Tang24} suggests that boil-off is the dominant mass loss mechanism shaping the small planet distribution, while photoevaporation deepens the radius gap and shifts the radius cutoff by affecting only 10\% of the total mass loss (a few percent drop in H/He mass fraction). From an evolution perspective, we find a similar trend in low-mass super-puffs: as shown in Figure \ref{tracks}, the evolution tracks of Kepler-51 planets, Kepler-47c, and Kepler-87c indicate that boil-off accounts for nearly 80-90\% of the total mass loss (usually reducing the envelope mass fraction from $\sim30\%$ to 10-20\% ) within the first 1–2 Myr, whereas photoevaporation contributes only the final 10-20\% of the total mass loss, equivalent to a 1-2\% drop in envelope mass fraction, over longer timescales. Note that at a given mass, increasing the incident flux leads to more rapid mass loss during the boil-off phase, making the mass loss contrast more pronounced. 

This conclusion holds for low-mass planets around G-type stars, while the effects of stellar type and rock/iron core mass are discussed below. 

Since boil-off is more sensitive to the rock/iron core mass than photoevaporation (see Section \ref{res:photo}), the relative role of photoevaporation increases for planets with higher core masses, which can retain their envelopes through boil-off even under stronger irradiation. This effect is evident in HIP 67522b, where its high mass ($>10 M_\oplus$) and intense irradiation ($>300 F_\oplus$) make photoevaporation more significant. 

In contrast, for Kepler-79d, photoevaporation dominates over boil-off due to both its high mass and host star type. F-type stars experience greater radius inflation on Gyr timescales than G-type stars. As a result, at the same present-day stellar luminosity, a planet around an F-type star would have received a lower bolometric flux during the boil-off phase, preserving a more substantial envelope and more inflated radius after boil-off and making photoevaporation more effective. Additionally, Kepler-79d’s assumed Earth-like bond albedo reduces the wind temperature, potentially limiting boil-off mass loss. This effect, which may be influenced by dusty outflows, warrants further study.

Although we do not include planets around stars of other spectral types, our analysis suggests that planetary evolution around early K-type stars resembles that around G-type stars, although with slightly stronger photoevaporation due to their higher XUV-to-bolometric luminosity ratio \citep{Jackson12}. However, for late-K and M-dwarfs, their prolonged pre-main-sequence contraction phase drastically alters planetary evolution: during the boil-off phase, planets around these stars experience bolometric fluxes over 100 times higher than their present-day values (assuming 5 Gyr). Moreover, at a given bolometric flux, the XUV flux is $\sim$10 times higher than that for planets around F- and G-type stars. These factors likely make both mass loss mechanisms important for planets around late-type stars, although boil-off can completely strip a planet before photoevaporation kicks off. Verification in future work is needed.

Based on these trends, we predict that within the parameter space we explored ($\leq$ 1 AU), super-puffs are more likely to occur around F- and G-type stars, while their H/He retention around M-dwarfs is challenging. Instead, we expect a large population of terrestrial and envelope-poor planets around M-dwarfs. At larger orbital separations (well beyond 1 AU), lower-density planets may form around M-dwarfs without being stripped by boil-off, but this remains observationally inaccessible, if they stay in-situ, and requires further study to assess consistency with dynamical models, if they migrate inward after boil-off.

\begin{figure}
\centering
\includegraphics[width=0.5\textwidth]{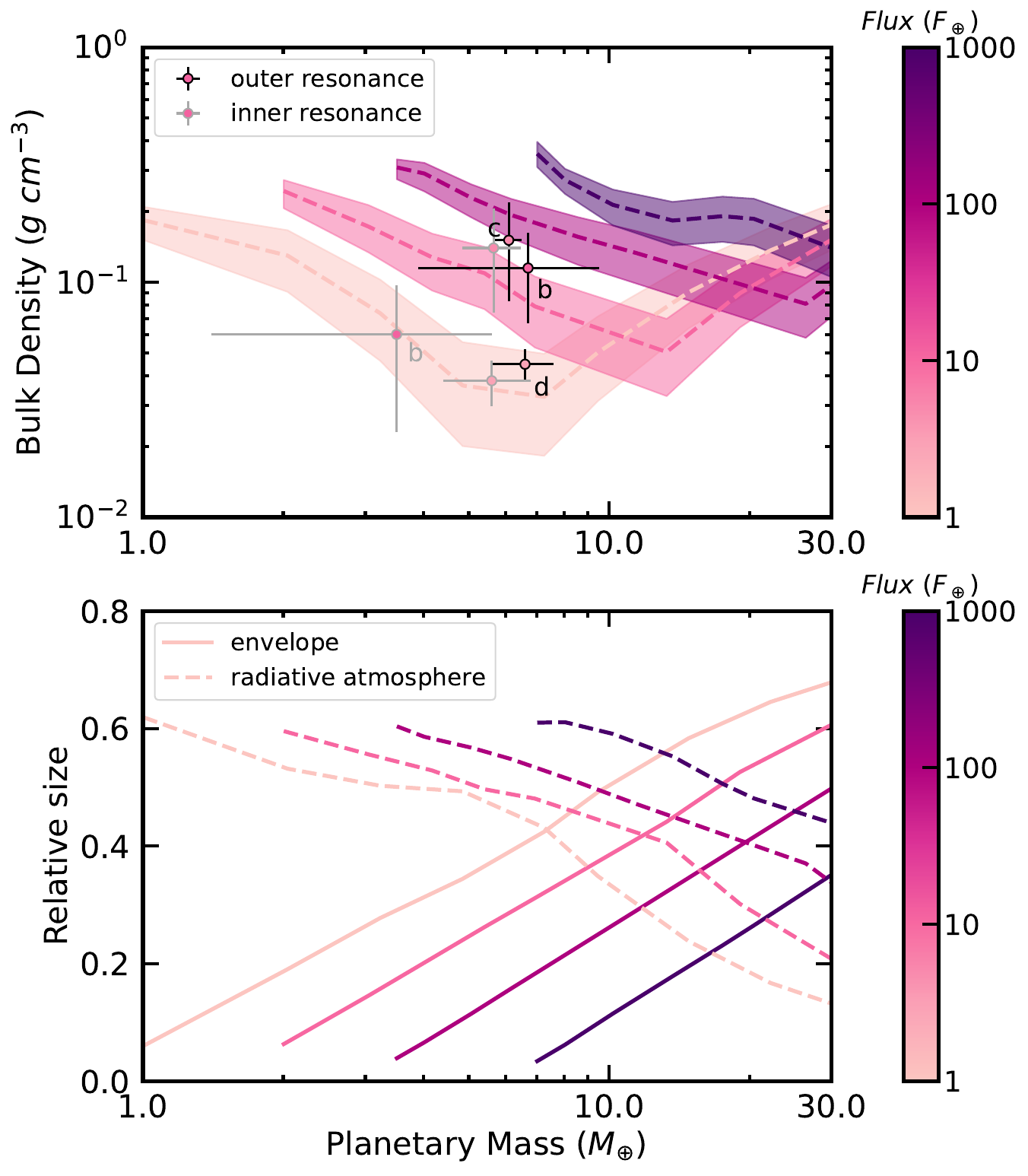} 
\caption{ The top panel shows the bulk density as a function of planetary mass (x-axis), with bolometric flux color-coded. For comparison, we overlay two sets of observations of the Kepler-51 planets, one from inner resonance (gray circles) and the other from outer resonance (black circles). The bottom panel presents the relative physical sizes of the envelope (solid) and radiative atmosphere (dashed), both normalized by the total planetary radius. This analysis assumes a hazy atmosphere with a transit pressure at 1 nbar, which is found to be likely for Kepler-51d in Figure \ref{mrk}.
}
\label{density}
\end{figure}

Given the dominant role of boil-off in super-puff mass loss, we isolate its effects by switching off photoevaporation and analyzing how super-puff properties are constrained in mass-radius-flux parameter space. The top panel of Figure \ref{density} shows the bulk density as a function of planetary mass, using Kepler-51 as a representative G-type host. Since these models constrain the maximum radius permitted by boil-off, the plotted bulk density represents the lowest possible value for each radius and incident flux. A higher bulk density formed from the gas-poor formation scenario \citep{Lee14} is theoretically possible but is beyond the scope of our study. Notably, the mass measurement assuming the inner resonance in \citet{Masuda24} is inconsistent with our model, as it implies a bulk density lower than the theoretical minimum. In contrast, the mass from the outer resonance (see the caption of Table \ref{table:para}) matches well. At its flux level (2.77$F_\oplus$), Kepler-51d represents the lowest bulk density achievable among planets with different rock/iron core masses (see the discussion below). Planets with even smaller bulk density than that of Kepler-51d is theoretically possible at lower fluxes.

Along an iso-flux curve, the lowest density occurs at the mass threshold where a planet is no longer susceptible to boil-off in its early evolution. As seen in the bottom panel of Figure \ref{density}, this minimum corresponds to the point where the envelope size becomes comparable to the thickness of the radiative atmosphere (solid and dashed curves, respectively). Less irradiated planets exhibit lower minimum bulk densities, as they can retain more H/He post-boil-off at low masses. The corresponding mass threshold also decreases with decreasing flux. These trends, derived for 1 nbar and specific ages, remain valid across different pressure levels and ages.

This analysis reveals that a super-puff requires both a substantial envelope and an extended radiative atmosphere, with the two components balancing at intermediate masses leading to their low bulk density. At low masses, planets experience a stronger variable gravity effect that inflates the radiative atmosphere but lose most of their envelope to boil-off. At higher masses, the envelope size is sufficient, but stronger gravity compresses the radiative atmosphere.

In conclusion, super-puff planets most commonly occur at low-to-moderate incident fluxes ($\lesssim 30 F_\oplus$) and intermediate masses (5–12 $M_\oplus$), depending on the flux level. When photoevaporation is considered, a low flux is necessary to preserve the super-puff characteristics over Gyr timescales. While this discussion focuses on G-type stars, it may extend to F-type and early-K stars, where super-puffs could form at higher and lower fluxes, respectively. However, such planets are unlikely around M-dwarfs.

\subsection{Constraining the Atmospheric Metallicity}
\label{res:metallicity}
\begin{figure}
\centering
\includegraphics[width=0.5\textwidth]{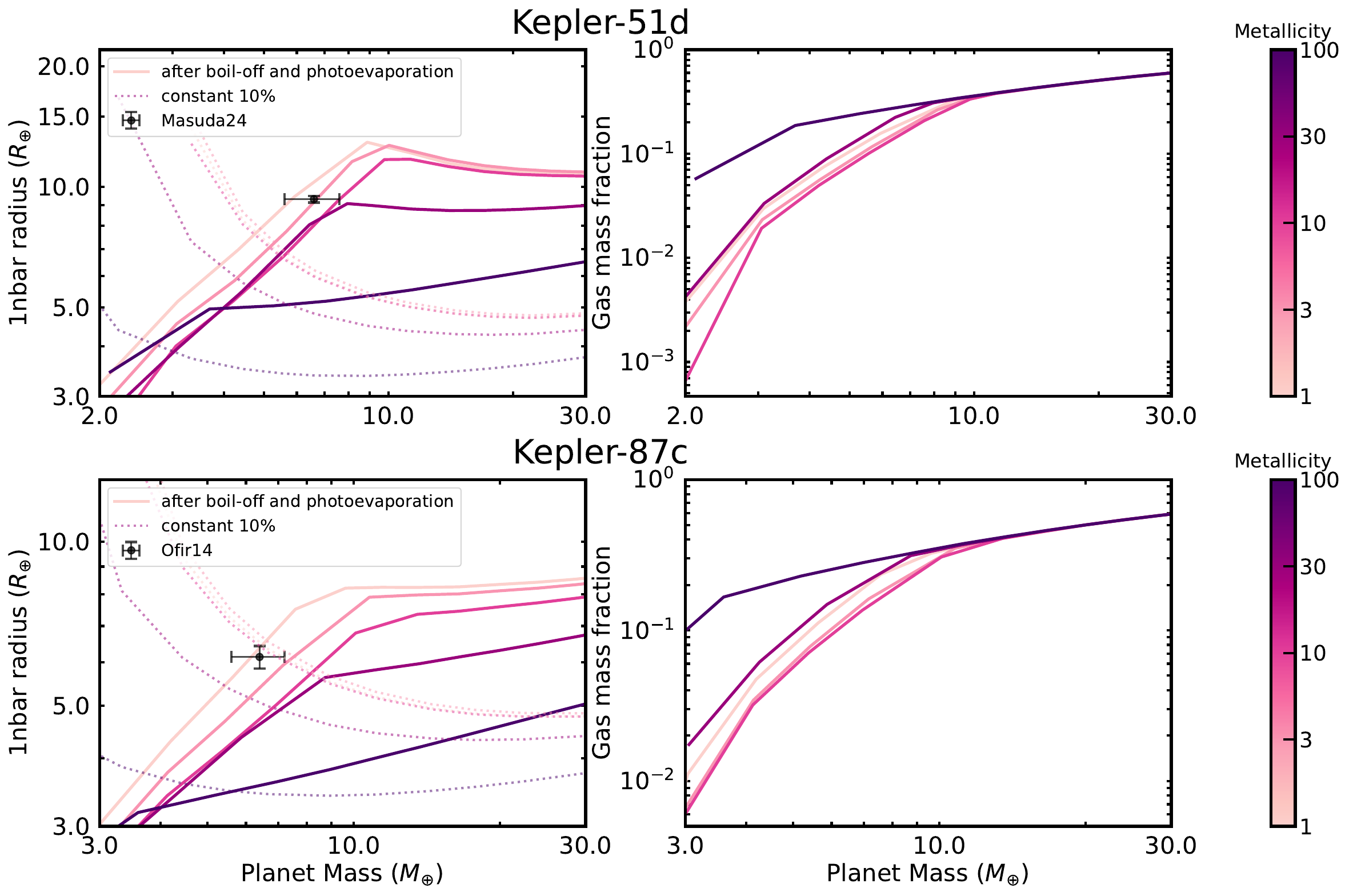} 
\caption{ M-R curves (left panel) and gas mass fraction as a function of planetary mass (right panel), with varied metallicity (color-coded) ranging from 1 to 100 times the solar value. Solid curves include both boil-off and photoevaporation, similar to Figure \ref{mrs}, while dotted curves assume constant compositions for comparison purpose. System ages are set to the observational lower bounds: 200 Myr for Kepler-51d and 5.68 Gyr for Kepler-87c. The results indicate that super-puffs likely have low atmospheric metallicity. See the main text for further discussion.
}
\label{mrm}
\end{figure}

While the previous sections focused on super-puff modeling with an atmospheric metallicity of one times solar, here we explore a broader range of metallicities and constrain their values for super-puffs in the context of our model. As described in Section \ref{subsec:photo}, we do not fully include metals in our outflow models but rather adjust the mean molecular weight of the outflowing H/He gas. In Figure \ref{mrm}, we examine metallicities from 1 to 100 times solar (color-coded) for Kepler-51d and Kepler-87c, which are representative of young and old super-puffs with hazy atmospheres. The left panels show the M-R relationship while the right panels present the total gas mass fraction as a function of planetary mass. 

To place an upper limit on atmospheric metallicity, we present the transit radii at the lower-bound system ages for each planet. The transit pressure level is set to 1 nbar, following the results discussed in Section \ref{res:app} and Figure \ref{tracks}. The solid M-R curves are constructed similarly to those in Figure \ref{mrk}, incorporating both boil-off and photoevaporation. For comparison, we also include iso-composition M-R curves (dotted), commonly used in other evolution studies, assuming a constant gas mass fraction of 10\%.

We identify five key effects of increasing atmospheric metallicity:
\begin{itemize}
    \item Density effect: higher envelope density reduces the envelope size, suppressing both photevaporation and boil-off.
    \item Entropy effect: slower intrinsic cooling leads to a more inflated envelope, promoting both mass losses.
    \item Scale height effect: a lower atmospheric scale height decreases the radiative atmospheric radius, reducing both mass losses.
    \item Particle weight effect: heavier outflow particles reduce boil-off \citep{Tang24}.
    \item Cooling effect: additional metal line cooling and enhanced electron number density reduce photoevaporation rates. However, this effect is minimal in super-puffs and is excluded from our model (see Section \ref{subsec:photo}).
\end{itemize}

\citet{Tang25} find that the density and entropy effects cancel out at 50 times solar metallicity. Expanding on this, we find that at low metallicities, the entropy effect slightly exceeds the density effect, such that increasing metallicity leads to greater boil-off mass loss (see right panels), which in turn reduces the envelope radius. However, above 10 times solar metallicity, the trend reverses: the density effect becomes dominant, leading to suppression of boil-off as metallicity increases. A similar reduction in mass loss occurs under photoevaporation, as the raising mean molecular weight strengthens the scale height effect. These effects lead to a smaller transit radius as metallicity increases. At metallicities exceeding 50 times solar, the density effect may become dominant over the entropy effect, depending on the envelope mass. Nevertheless, for sub-Neptune planets, the combined influence of the entropy and density effects remains limited, causing at most a 20\% difference in envelope radius. This effect is more likely to be significant in Saturn- and Jupiter-mass planets.


In addition, metallicity affects the radius evolution via mass loss, primarily through both the particle weight and scale height effects. These effects are more pronounced during boil-off, which dominates over photoevaporation. Importantly, the positive dependence of the mass loss rate on $\mu_s$ in the TEMP regime (Eq. \ref{mdot-scaling}) counteracts the scale height effect (through $R_{\rm{base}}$), making TEMP less sensitive to metallicity than the energy-limited phase.

At elevated metallicity ($\geq$10 times solar), the scale height effect typically leads to a greater radius reduction than the combined density and entropy effect, as seen in the iso-composition curves (dotted) in Figure \ref{mrm}. This effect is amplified at lower gravity. As a consequence, we find that the transit radius in the presence of mass loss (solid) generally decreases monotonically with increasing metallicity at a given mass. However, at very high metallicities $\gtrsim$50 times solar and for low-mass planets, the transit radius increases with metallicity due to the higher envelope mass that planets can retain following boil-off. Given this behavior and the observational uncertainties in mass and radius (black error bars), in the context of our model, we constrain the metallicities of Kepler-51d and Kepler-87c to no more than 10 times solar and 3 times solar, respectively (see left panels of Figure \ref{mrm}).

In conclusion, we suggest that super-puffs have relatively low atmospheric metallicities, likely below 10 times solar, with a corresponding mean molecular weight of 2.62, which is an essential condition for explaining their extraordinarily large radii. Since a hazy atmosphere requires sufficient metals as building blocks, further studies are needed to assess the consistency of these findings with haze formation models.

\section{Discussion} \label{sec:disc}
\subsection{Stellar Wind} \label{disc:wind}
The stellar wind is neglected in the photoevaporation model. We evaluate its strength through the total pressure $P_*$, which is the sum of thermal pressure $n_*kT_*$, magnetic pressure $B_*^2/(8\pi)$, and ram pressure $n_*m_Hv_*^2$. For a hot Jupiter subjected to the solar wind, this pressure is estimated to be on the order of 10 picobar \citep{RMC09}. For a close-in young super-puff, a significantly stronger stellar wind is expected due to enhanced stellar activity and a low photoionization base pressure resulting from its low gravity. Among all model planets, the highest stellar wind pressure reaches approximately 0.1 nbar (calculated from the stellar wind model introduced in Section \ref{subsec:stellar}). If this pressure is comparable to or exceeds the photoionization base pressure, it could challenge our assumption by confining the wind, reducing the mass loss rate, and altering the wind structure by influencing the neutral column density.

For all models in this study, we find that the photoionization base remains well below the stellar wind front, with $P_*<0.01P_{\rm{base}}$. This is because stellar wind strength increases with incident flux, and highly irradiated planets also experience stronger boil-off, which raises their gravity and leads to a correspondingly higher base pressure (see Section \ref{res:photo}). Therefore, stellar wind effects are not relevant to super-puffs.

\subsection{Tidal Heating} \label{disc:tides}
\begin{figure}
\centering
\includegraphics[width=0.5\textwidth]{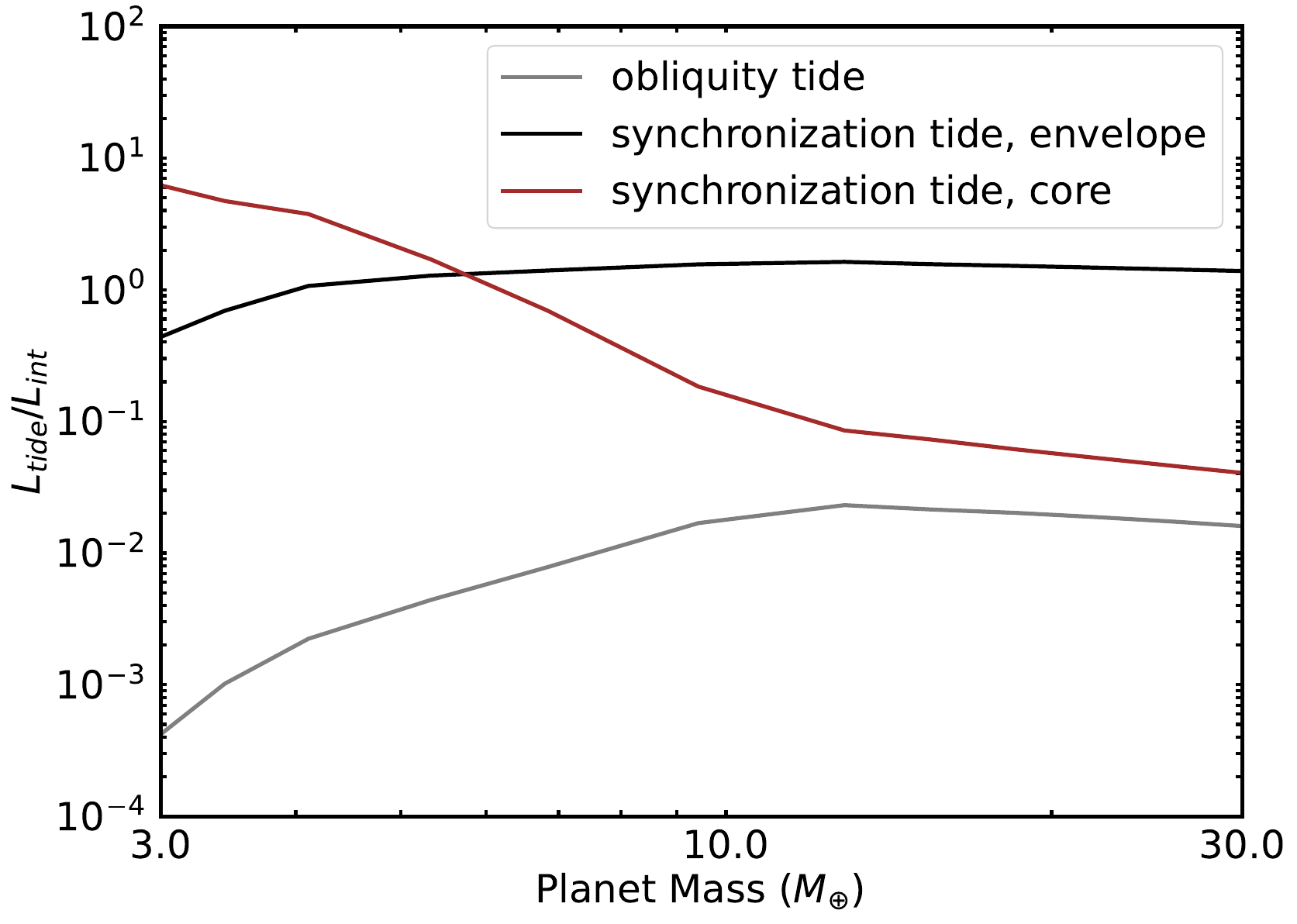} 
\caption{ Tidal heating normalized by the total interior cooling as a function of planet mass for Kepler-51b. Three types of tides are considered here at the observed age: obliquity tide (gray), synchronization tide in the envelope (black) and synchronization tide in the rock/iron core (red).
}
\label{tides}
\end{figure}

We assess tidal heating using the framework developed by \citet{Leconte10}. We neglect the contribution from orbital eccentricity $e$, given the low $e$ values typically observed in Kepler multi-planet systems \citep{Vaneylen15}. We consider two primary types of tidal heating: (1) Obliquity tides, which arise when a planet’s rotation has evolved into an equilibrium state with non-zero obliquity ($\epsilon$). 
(2) Synchronization tides, which occur when a planet’s initial rotation rate exceeds its orbital frequency, gradually leading to rotational synchronization. In Figure \ref{tides}, we present the effects of obliquity tides (gray) and synchronization tides deposited in the envelope (black) and core (red) as a function of planetary mass, for Kepler-51b, which among super-puffs has a shorter orbital separation and is more susceptible to the tidal heating. The tidal power $L_{\rm{tide}}$ is normalized by the intrinsic luminosity $L_{\rm{int}}$ to quantify its relative importance for thermal evolution.

We find that obliquity tides are negligible in super-puffs, primarily because these planets are expected to receive low incident flux in order to maintain their large physical sizes, and tidal power declines steeply (as a sixth power) with increasing separation. We estimate that tidal heating contributes less than 1\% of the total cooling luminosity for a typical super-puff. This result contrasts with \citet{Millholland19}, which suggested that obliquity tides are significant for super-puffs. 

The discrepancy arises from both updated planetary parameters and the choice of planetary radius used in estimating tidal power. Our model indicates that the planets in the Kepler-27, Kepler-31, and Kepler-79 systems, when modeled with the parameters adopted in \citet{Millholland19}, would not survive a boil-off phase due to their low masses and high incident fluxes. These conditions lead to an overestimation of tidal heating in \citet{Millholland19}. Moreover, because a large fraction of a super-puff’s radius comes from its extended radiative atmosphere (see Section \ref{res:implication}), which has negligible mass and thus does not contribute significantly to tidal dissipation, we argue that the envelope radius, rather than the optical transit radius, is more appropriate for tidal heating calculations. Finally, given their wide orbital separations and low masses compared to hot-Jupiters, super-puffs have synchronization timescales that are longer than Gyr timescales. We therefore conclude that these planets are likely not in the equilibrium rotation regime.

The dominant source of tidal heating arises from rotation-orbit interactions during the synchronization process. In this regime, the planetary rotation rate $\omega$ enhances the tidal power by a factor of $(\omega/n - 1)^2$ relative to obliquity tides, where $n$ is the orbital mean motion. We find that synchronization tides can potentially generate heating comparable to the total cooling luminosity of super-puffs with smaller orbital separations (e.g. Kepler-79d and Kepler-51b). For these planets, we assume a rotation rate of 10\% of the break-up spin in Figure \ref{tides}, which is smaller than the $\sim$30\% observed for Kepler-51d \citep{Liu25} due to the shorter synchronization timescales. Such heating may offer an alternative explanation for their large physical sizes. In contrast, tidal heating is limited in Kepler-51d, which has a larger orbital separation. 

Given that the rock/iron core remains largely molten in sub-Neptunes \citep{Tang25}, we adopt a tidal quality factor of $Q_c = 10^4$ for the rock/iron core \citep{Zahnle15} and $Q_{\rm{env}} = 10^5$ for the H/He envelope \citep{Goldreich66}. Under these assumptions, we find that atmospheric tides in the envelope dominate over those in the core in intermediate-mass (most common for super-puffs) and higher-mass planets, because the envelope constitutes a large fraction of their physical sizes. For low-mass planets, core tides may dominate because boil-off significantly reduces the envelope radius. However, due to uncertainties in the planetary rotation rate and tidal dissipation coefficients, this analysis remains speculative and warrants further investigation.

\subsection{Future Work}
Future extensions of this framework include: (1) Conducting a comprehensive population study to compare our model predictions with the observed small-planet radius distribution, which will provide insights into the radius gap and radius cutoff, further validating our model. (2) Expanding the analysis to include a broader range of stellar spectral types, particularly M-dwarfs, as planetary evolution in these systems can differ significantly. (3) Incorporating the evolution of envelope-free planets, which commonly appear near or within the radius gap. This addition will be crucial for understanding the observed gap’s width and depth.

\section{Conclusions} 
We propose that super-puffs initially form with a substantial H/He envelope ($\sim 30\%$), likely in-situ losing a moderate fraction of this mass through a combination of boil-off and photoevaporation, with boil-off dominating the mass loss. This evolution ultimately leaves them with an envelope comprising 10–20\% of their total mass. The very good agreement between our predicted radii and the observed masses and ages of several super-puffs and young planets supports this formation scenario.

Super-puffs occupy an intermediate mass range and experience low to moderate stellar irradiation. These conditions allow them to retain a thick envelope after boil-off while maintaining a large atmospheric scale height, which inflates their apparent radius. We find that low atmospheric metallicity plays a key role in sustaining a thick radiative atmosphere by enhancing the scale height. Other metallicity effects are generally minor, except for its influence on mean molecular weight of the boil-off outflow, which impacts mass loss. At a given stellar insolation, super-puffs are more likely to form around G- and F-type stars than M dwarfs and late-K-type stars, as they undergo less extreme boil-off and photoevaporation.

Our model predicts that Kepler-51d, Kepler-87c, and HIP 67522 b likely host hazy atmospheres, while V1298 Tau b is expected to have a clear atmosphere. We recommend follow-up spectroscopic observations of Kepler-87c to test this prediction. For other planets, significant uncertainties in mass measurements prevent a definitive determination of their transit pressure levels. However, our model finds that the pressure levels for those plants that may host hazy atmospheres fall within the range of 1–100 nbar (see Kepler-51d in Figure \ref{mrk} as an example), consistent with photochemical modeling.

We identify a distinct mass loss regime, which we termed thermal-energy-mediated photoevaporation (TEMP), for low-gravity planets, where EUV energy input is nearly completely converted into temperature-dependent thermal energy rather than gravitational potential energy, which depends on planetary mass and radius. The thermal-energy-mediated regime commonly occurs in low-mass ($\lesssim 13 M_\oplus$) sub-Neptunes that have undergone a boil-off phase. We derive an analytical scaling relation for this new regime and establish transition criteria. We find that recombination-limited mass loss and X-ray-driven escape do not happen in low-mass, low-gravity planets, nor is thermal-energy-mediated photoevaporation relevant for high-mass planets.

Previous models systematically overestimated the role of photoevaporation in low-mass planets due to two main factors: (1)the neglect of the boil-off phase, which leads to an overestimation of the initial mass fraction and thus $R_{\rm{base}}$, and (2) the analytical generalization from hot Jupiter models that energy sinks into thermal energy are negligible does not hold for sub-Neptunes. To address these uncertainties, we present analytical formulations for the thermal-energy-mediated photoevaporation (Eq. \ref{mdot-scaling} and \ref{thermal-limited-mdot}). Together with the physically motivated initial conditions derived from boil-off modeling (Table 5 of \citet{Tang25}), these offer a readily applicable improvement to the accuracy of other sub-Neptune evolution frameworks.

Additional uncertainty in previous mass loss estimates arises from the simplified structure and evolution framework \citep{Tang25}, which significantly impacts $R_{\rm{base}}$. 

To refine our understanding of these processes and further validate our framework, we propose a comprehensive population study incorporating a wider spectrum of stellar types.

\bibliography{reference}{}
\bibliographystyle{aasjournal}

\appendix
\restartappendixnumbering
\section{Adjustment of mean molecular weight in modeling photoevaporative outflows} \label{app:mmw}
In this work, we account for the effect of metallicity on photoevaporative outflows by adjusting the mean molecular weight of the wind. In this appendix, we discuss the potential uncertainties associated with this simplification.

First, the metal species may reduce the photoevaporation mass-loss rate by cooling the wind through metal line radiation. However, due to the low wind temperature, resulting from the wide orbital separations of super-puffs, and the correspondingly low ionization fraction, this cooling  is expected to be inefficient (see the bottom panel of Figure \ref{energy}, which shows Ly-$\alpha$ cooling).

\begin{figure}
\centering
\includegraphics[width=0.45\textwidth]{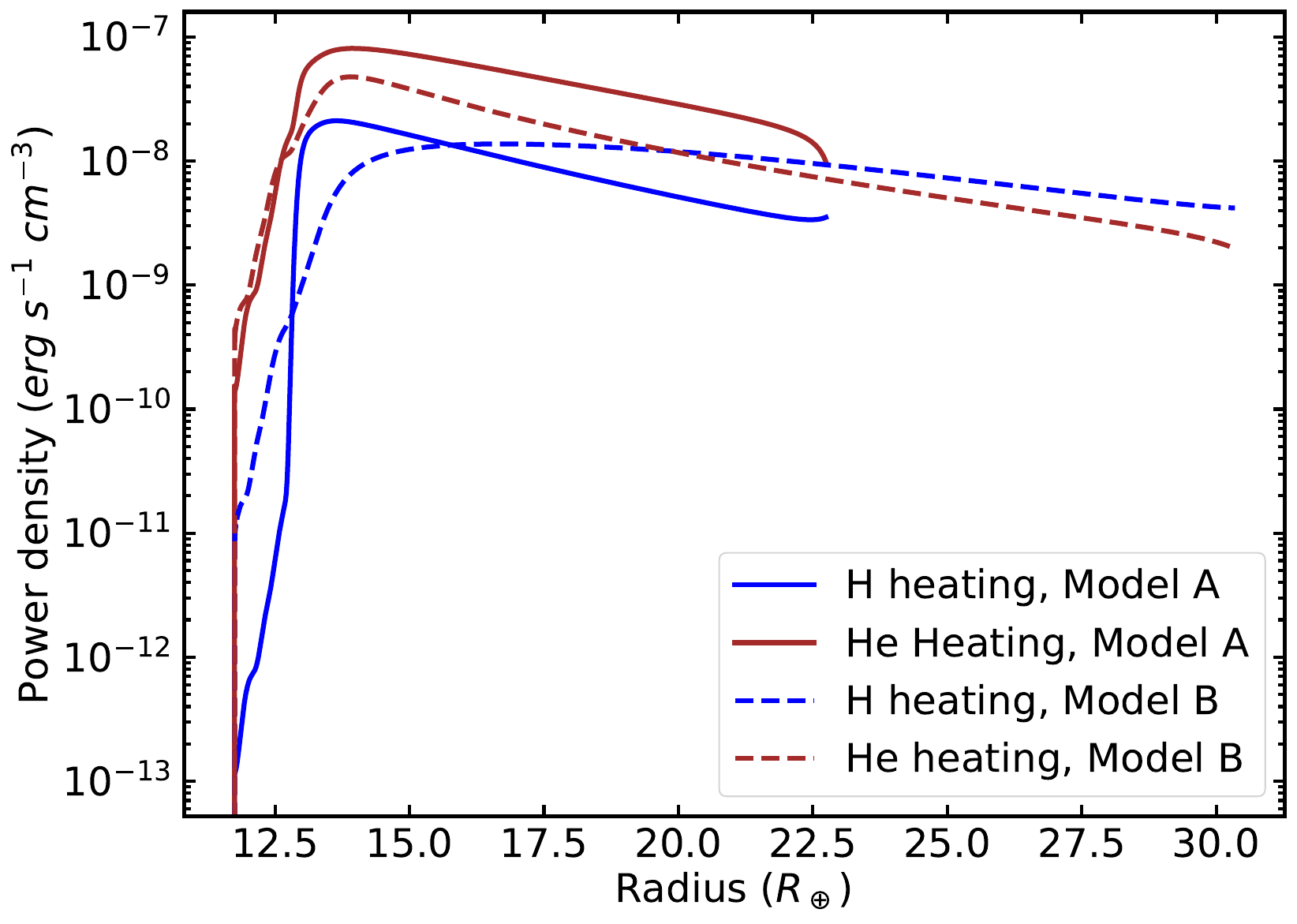} 
\caption{ Comparison of heating profiles from Model A (solid) and Model B (dashed), both with the same mean molecular weight, for Kepler-51d at 100× solar metallicity. Model A achieves the target molecular weight by adjusting the H–He mass ratio, while Model B retains the unadjusted solar H–He mass ratio and instead varies the hydrogen atomic mass. 
}
\label{heating}
\end{figure}

Second, hydrogen, helium and metal species absorb XUV energy at different rates and wavelengths. Neglecting metal species and adjusting the H–He ratio may alter the photoionization heating profile, potentially affecting the mass-loss rate. Super-puffs may be particularly susceptible to this effect, as Section \ref{res:photo} suggests that their XUV absorption regions are significantly extended.  To assess this uncertainty, we explore an alternative method for adjusting the mean molecular weight by varying the hydrogen atomic mass (Model B) and compare it to our default approach of varying the H-He mass ratio (Model A). Figure \ref{heating} presents the heating power density of hydrogen (blue) and helium (pink), demonstrating that the spectrum dependence induces variations in the heating profile for super-puffs. Nevertheless, our results indicate that the spectrum-dependent uncertainty in the mass-loss rate is typically within approximately 10\%, even at high metallicities ($1.45\times10^{11} g\ s^{-1}$ for Model A vs. $1.63\times10^{11} g\ s^{-1}$ for Model B).

\section{Analytical fits of wind parameters} 
\label{app:fit}
In Section \ref{res:photo}, we present analytical fits to the wind parameters based on a set of 17 model planets spanning a broad range of planetary masses and bolometric fluxes.

\begin{figure}
\centering
\includegraphics[width=0.45\textwidth]{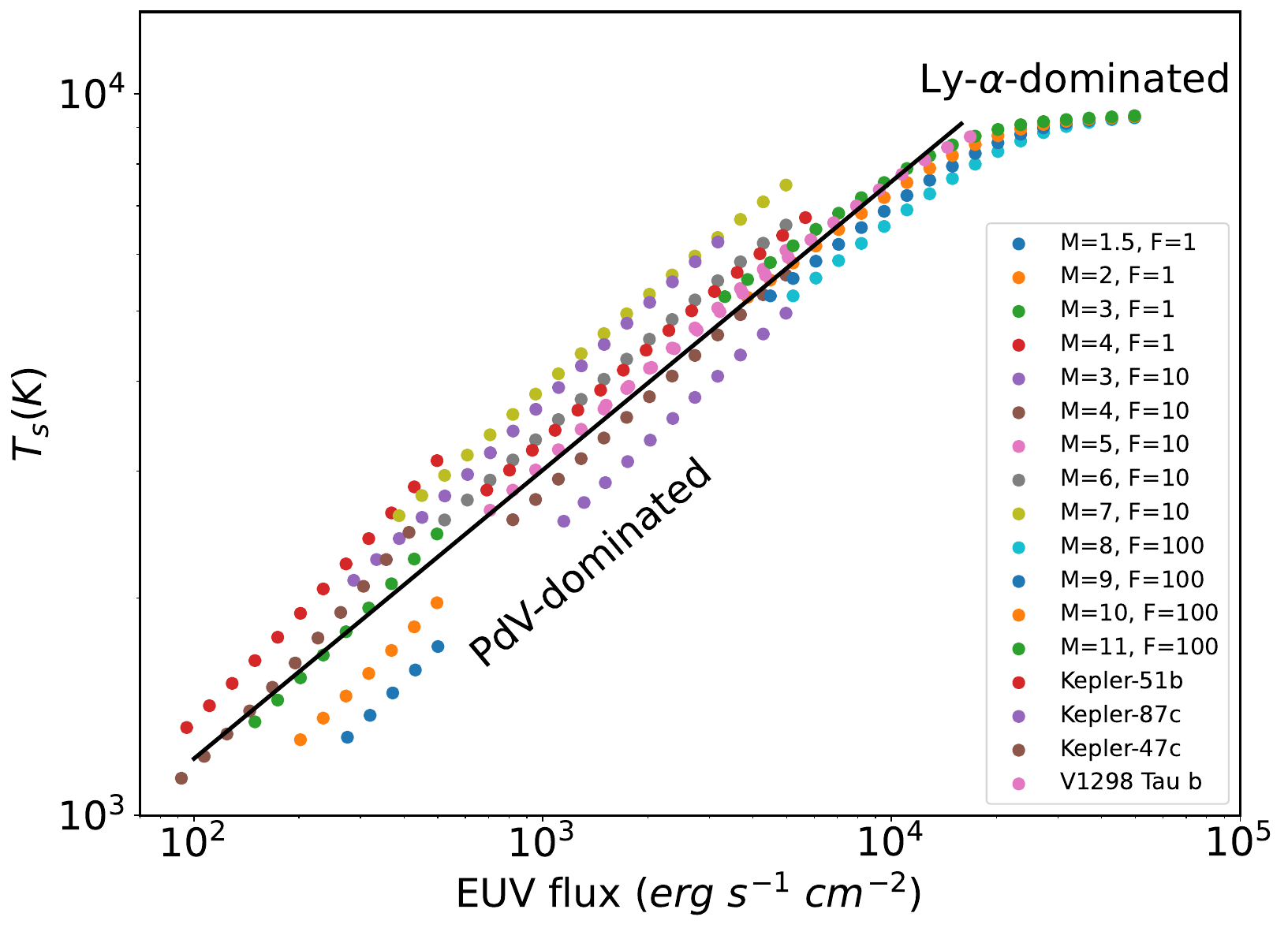} 
\caption{ Temperature at the sonic point as a function of the EUV flux. The analytical fit is shown in black and the data points are color-coded.
}
\label{Ts}
\end{figure}

In Figure \ref{Ts}, we show the temperature at the sonic point $T_s$, as a function of the EUV flux $F_{\rm{EUV}}$. For temperatures $\lesssim$10000 K, the outflow dominated by $PdV$ work, and Ly-$\alpha$ cooling (along with other line cooling) is negligible. In this regime, $T_s$ follows a power-law scaling with $F_{\rm{EUV}}$ (Eq. \ref{temperature-scaling}) along the evolutionary track of a given planet (from top right to bottom left). We find that the power-law slope remains consistent at 0.4 across all sample planets, while the coefficient shows a weak dependence on the rock/iron core mass, increasing slightly with heavier cores for a given bolometric flux.  Notably, for planets that have undergone boil-off, heavier cores correspond to lower gravities (see Figure \ref{density}). When $F_{\rm{EUV}}$ becomes sufficiently high, $T_s$ saturates at 10,000 K due to efficient Ly-$\alpha$ cooling.

\begin{figure}
\centering
\includegraphics[width=0.45\textwidth]{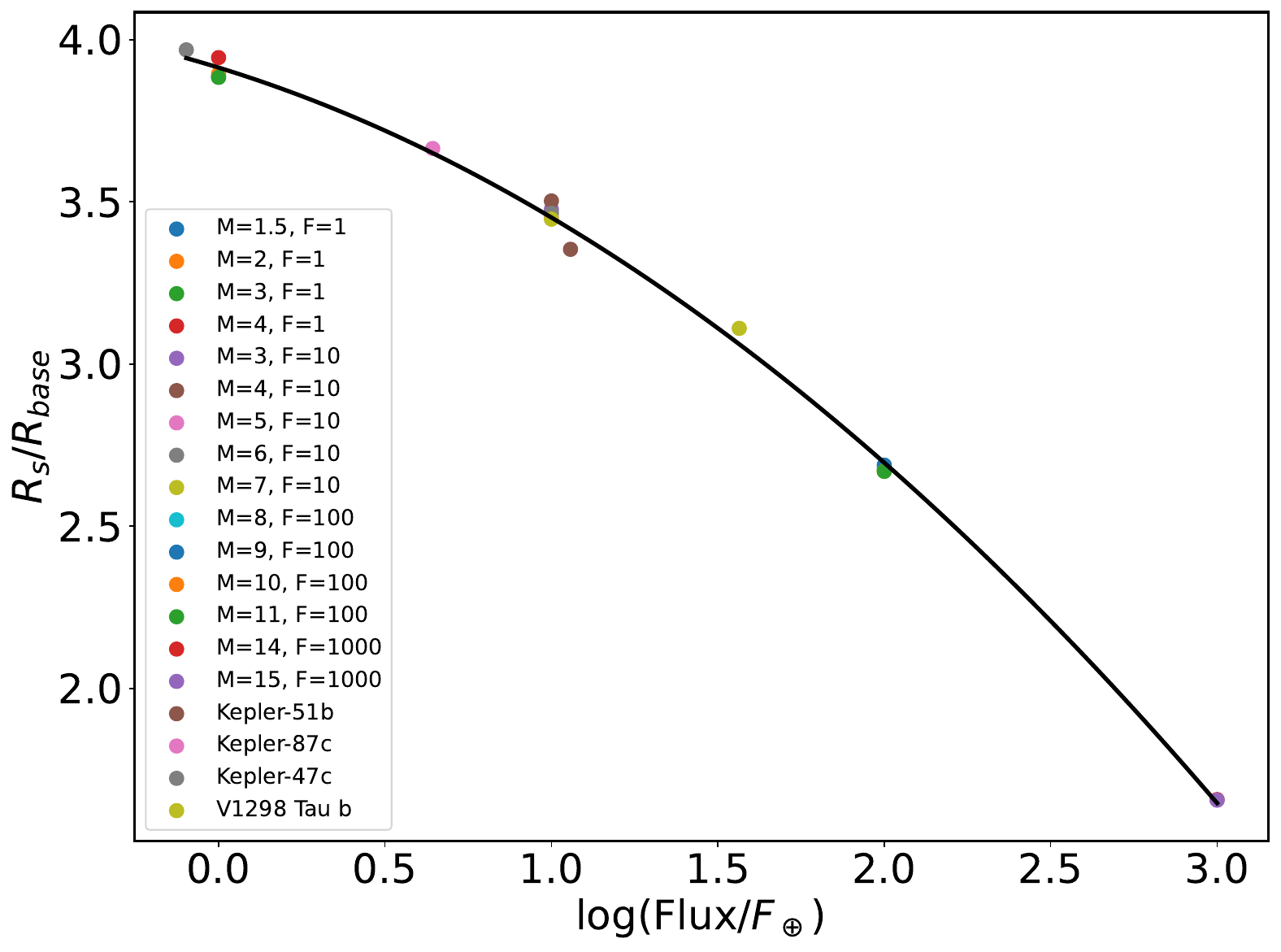} 
\caption{ The ratio of the sonic radius to the photoionization base radius at the transition between the two mass-loss regimes, as a function of the logarithm of bolometric flux. The analytical fit is shown in black and the data points are color-coded.
}
\label{rs}
\end{figure}

In Figure \ref{rs}, we present the radius ratio $R_s/R_{\rm{base}}$ at the transition between the thermal-energy-mediated and energy-limited regimes, plotted as a function of the logarithm of bolometric flux $F$.

\begin{figure}
\centering
\includegraphics[width=0.9\textwidth]{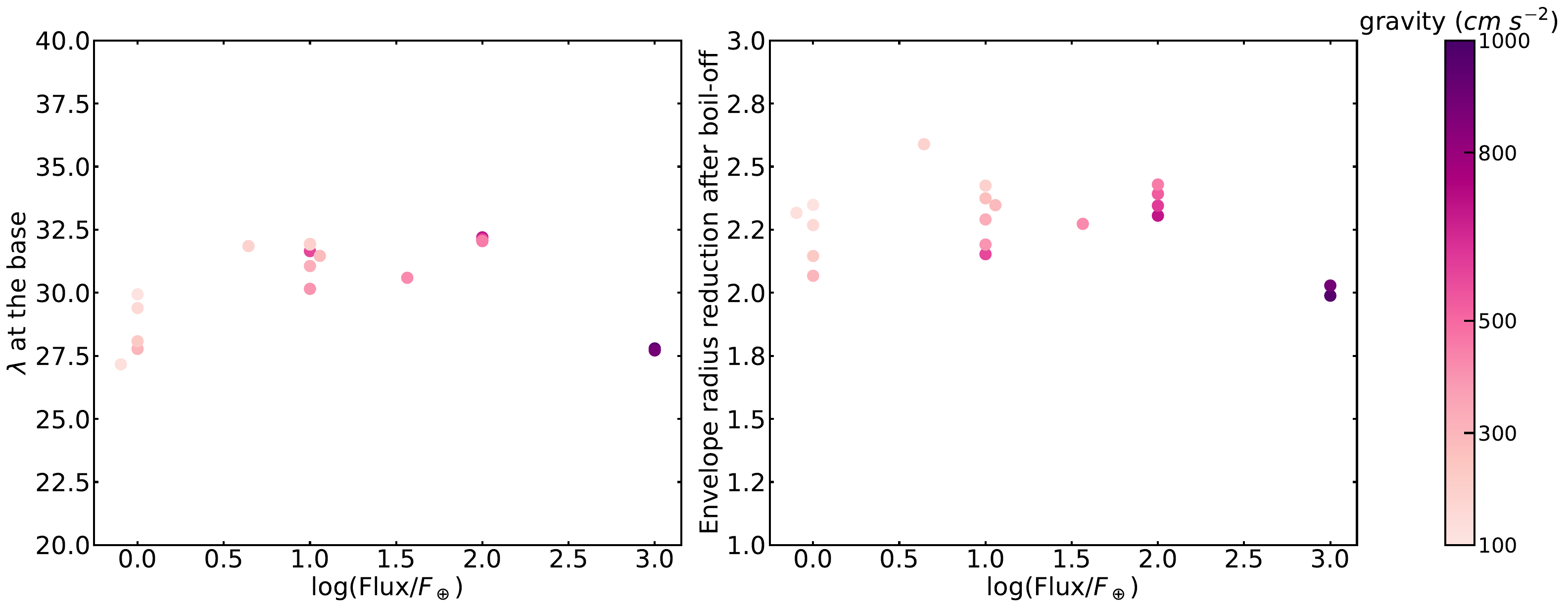} 
\caption{ Atmospheric escape parameter at the wind base as a function of flux and gravity (left) and envelope radius reduction following boil-off (right), for the same sample of planets shown in Figure \ref{Ts} and \ref{rs}.
}
\label{lambdas}
\end{figure}

The left panel of Figure \ref{lambdas} shows the atmospheric escape parameter $\lambda$ at the wind base as a function of incident bolometric flux (y-axis) and gravity (color-coded, defined at 20 mbar). It typically reaches a value of 30, only weakly dependent on gravity and flux levels, when the transition between the two mass loss regimes occurs. The right panel shows the ratio of the envelope radius between the end of boil-off and the mass-loss regime transition. The transition typically corresponds to a factor of 2 envelope size reduction since the end of boil-off.

\begin{figure}
\centering
\includegraphics[width=0.45\textwidth]{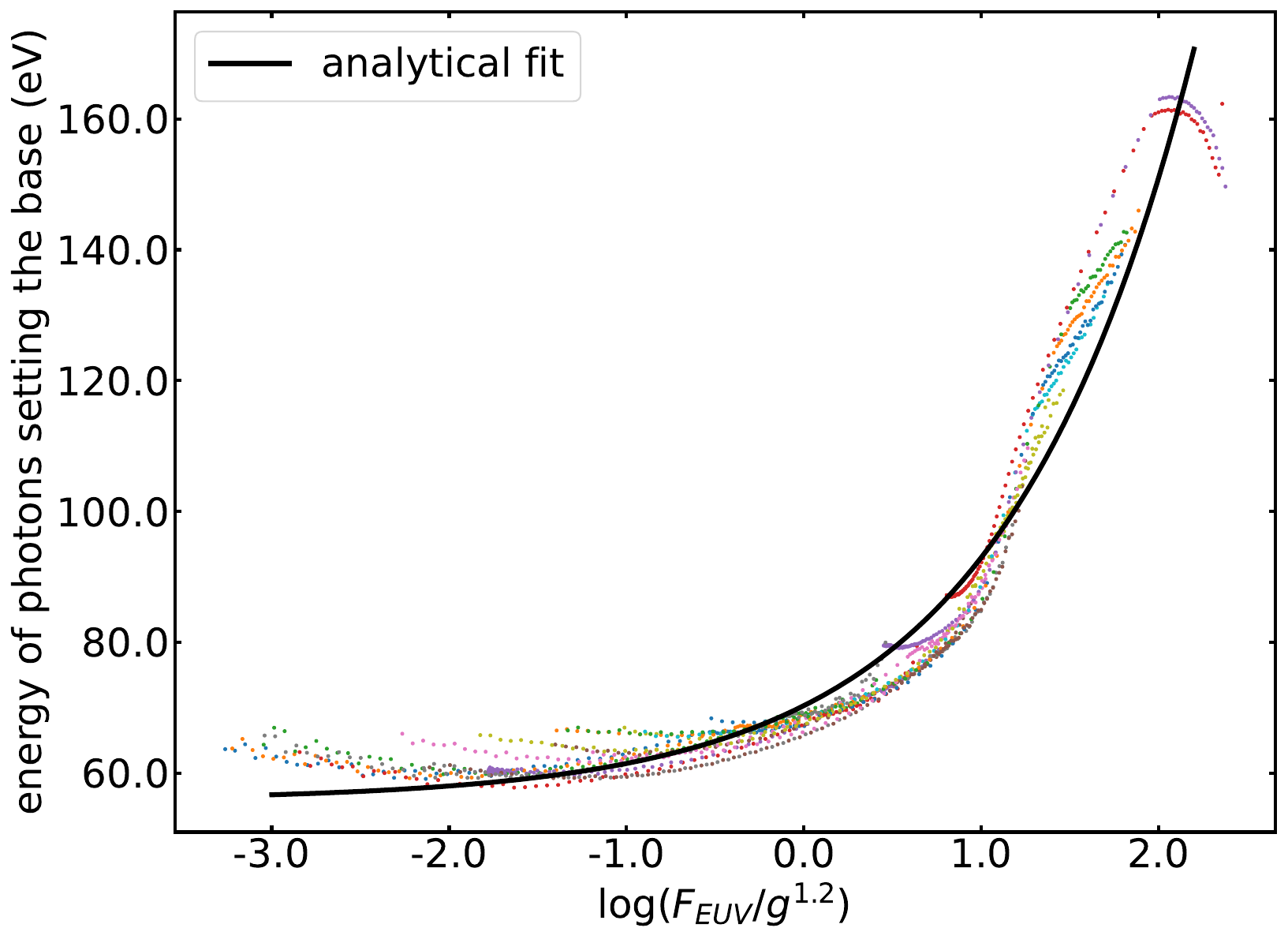} 
\caption{ Analytical fit of the characteristic photon energy that sets the wind base, for the same sample of planets shown in Figure \ref{Ts} and \ref{rs}.
}
\label{energy_fit}
\end{figure}

Additionally, Figure \ref{energy_fit} shows the characteristic photon energy that penetrates to the wind base, fitted as a function of $F_{\rm{EUV}}/g^{1.2}$. Notably, for planets (particularly those at late ages) in the energy-limited phase, this characteristic photon energy is roughly constant (see the lower left portion of the fitted curve). For reference, planets evolve along the color-coded tracks from the upper right to lower left along color-coded curves.

\end{document}